# A General Framework for Liquid Marbles

A. Gallo Jr.[1]*, F. Tavares[1], R. Das[1], H. Mishra[1]*

[1]Interfacial Lab (iLab), Water Desalination and Reuse Center (WDRC), Division of Biological and Environmental Sciences (BESE), King Abdullah University of Science and Technology (KAUST), Thuwal 23955-6900, Saudi Arabia.

*Himanshu.Mishra@Kaust.edu.sa; Adair.GalloJunior@Kaust.edu.sa

**Abstract**

Liquid marbles refer to liquid droplets that are covered with a layer of non-wetting particles. They are observed in nature and have practical significance. However, a generalized framework for analyzing liquid marbles as they inflate or deflate is unavailable. The present study fills this gap by developing an analytical framework based on liquid–particle and particle–particle interactions. We demonstrate that the potential final states of evaporating liquid marbles are characterized by one of the following: (I) constant surface area, (II) particle ejection, or (III) multilayering. Based on these insights, a single-parameter evaporation model for liquid marbles is developed. Model predictions are in excellent agreement with experimental evaporation data for water liquid marbles of particle sizes ranging from 7 nm to 300 μm (over four orders of magnitude) and chemical compositions ranging from hydrophilic to superhydrophobic. These findings lay the groundwork for the rational design of liquid marble applications.

**Teaser**
Underpinnings of the curious behaviors of liquid marbles—water drops covered with water-repelling particles—are revealed.

**Introduction**

Liquid marbles are commonly composed of water droplets and covered with a layer of hydrophobic particles or powders (*1-4*). This arrangement prevents direct contact between the liquid and underlying substrate. Thus, a liquid marble is a "non-wetting soft object" (*5*) that rolls and bounces like a marble when gently displaced. Curiously, aphids residing inside confined plant galls prepare liquid marbles to preempt life-threatening risks of getting wet by their own sugary secretions. To mitigate this, aphids coat the sticky secretions with wax particles to produce ~0.1-mm-diameter non-sticky marbles for waste disposal (*6, 7*). Aussillous and Quéré were the first to report liquid marbles in laboratories (*1*). Subsequently, many reports on the fundamental characteristics of liquid marbles have appeared such as on their evaporation (*8*), coalescence (*9*), physical partitioning (*3*), viscous dissipation during rolling (*1, 10*), and exposure to electromagnetic fields (*3*). Several potential applications of liquid marbles have also been explored such as for detecting water pollution (*11*), monitoring environmental gases (*12*) and interfacial reactions (*13*), bioreactors for blood typing (*14*), cell culturing and screening (*15-18*), polymerase chain reaction assay (*19*), electrochemistry (*20*), micellar self-assembly (*14, 15, 17, 19, 21, 22*), and magnetic translocation (*23-26*) among others (*2, 27-32*). Majority of these fundamental and applied studies evaluate liquid marbles thermo/electro/mechano-/magneto-statically. Despite the considerable interest and value of the above research, a unified framework for describing the mechanics of stressed liquid marbles, especially as they deflate, is unavailable. Analytical approaches for modeling the evaporation of sessile liquid marbles (*8, 33*) exploit empirical parameters that may not provide physical insights into the role of liquid–particle interfacial tension, particle surface roughness (*34*), particle–particle friction coefficient (*35*), and other attributes such as interfacial electrification (*36-39*).

Particles constituting the shell of a stationary spherical liquid marble experience coupled forces owing to weight ($F_w$), buoyancy ($F_b$), liquid–particle adhesion ($F_{adh}$), capillarity that exerts a compression force ($F_c$) on the particles, and interparticle friction ($F_{fr}$) that depends on $F_c$ and interparticle friction coefficient. Particles at the bottom also experience the weight of the drop, which pushes them inside the liquid. If an external stimulus, e.g., mechanical collision (*9*) or liquid withdrawal via evaporation (*8*), stresses the liquid marble, additional forces may appear. In practice, when the liquid is withdrawn from a liquid marble, drastically varying scenarios may occur including the physical distortion of liquid marbles such as buckling and crumpling (*8, 40, 41*), particle multilayering (*8, 33*), or even the ejection of some particles from the liquid marble into the air (*42, 43*). Even though the liquid–particle and interparticle forces majorly dictate the outcomes, a generalized framework for analyzing liquid marbles based on these forces is lacking. Herein, a complementary experiment and theory are combined to fill this gap. First, the study investigates the evaporation of formed liquid marbles using particles of sizes varying over four orders of magnitude (7 nm–300 µm) and chemical compositions ranging from hydrophilic to superhydrophobic (Table 1, Figs. 1–3). Then, the general framework for analyzing liquid marbles is presented, which considers forces generated through liquid–particle and particle–particle interactions as well as an ejection force resulting from liquid removal (Figs. 4–5). Finally, this study constructs a single-parameter evaporation model based on these insights to describe the potential final states of the variegated liquid marbles introduced above and detailed below (Figs. 6–7).

**Results**

*Liquid marble preparation and characterization*

This study uses batches of silica particles with a characteristic dimension, $d_p$, ranging from 7 nm to 300 µm, to disentangle the effects of particle size and roughness on liquid marble behavior (Figs. 1–2). Silica surfaces enabled the precise control of their chemical composition via silanization reactions (Figs. 1A–B, Table 1, Methods). Thus, the wettability of particles could be tuned from hydrophilic to hydrophobic by varying the length of alkyl chains, e.g., from octyl (C8) to octadecyl (C18), that were chemically grafted onto them. Glaco Mirror Coat™ was applied onto 128-µm-sized silica particles that were already coated with C18 to render them superhydrophobic at grain-level.

To create a liquid marble, a 10-µL droplet of water was placed on an ~2-mm-thick layer of hydrophobic particles on a glass slide. Then, the hydrophobic particles were gently poured (from top) over the drop; they initially slid down the drop's surface, and the particulate layer grew bottom–up and eventually covered the entire water droplet (Fig. 1C, Movie S1). The surface particle density, $\sigma$, which is defined as the mass of the particles divided by the liquid surface area, can be obtained by comparing the mass of the water droplet before and after creating the liquid marble.

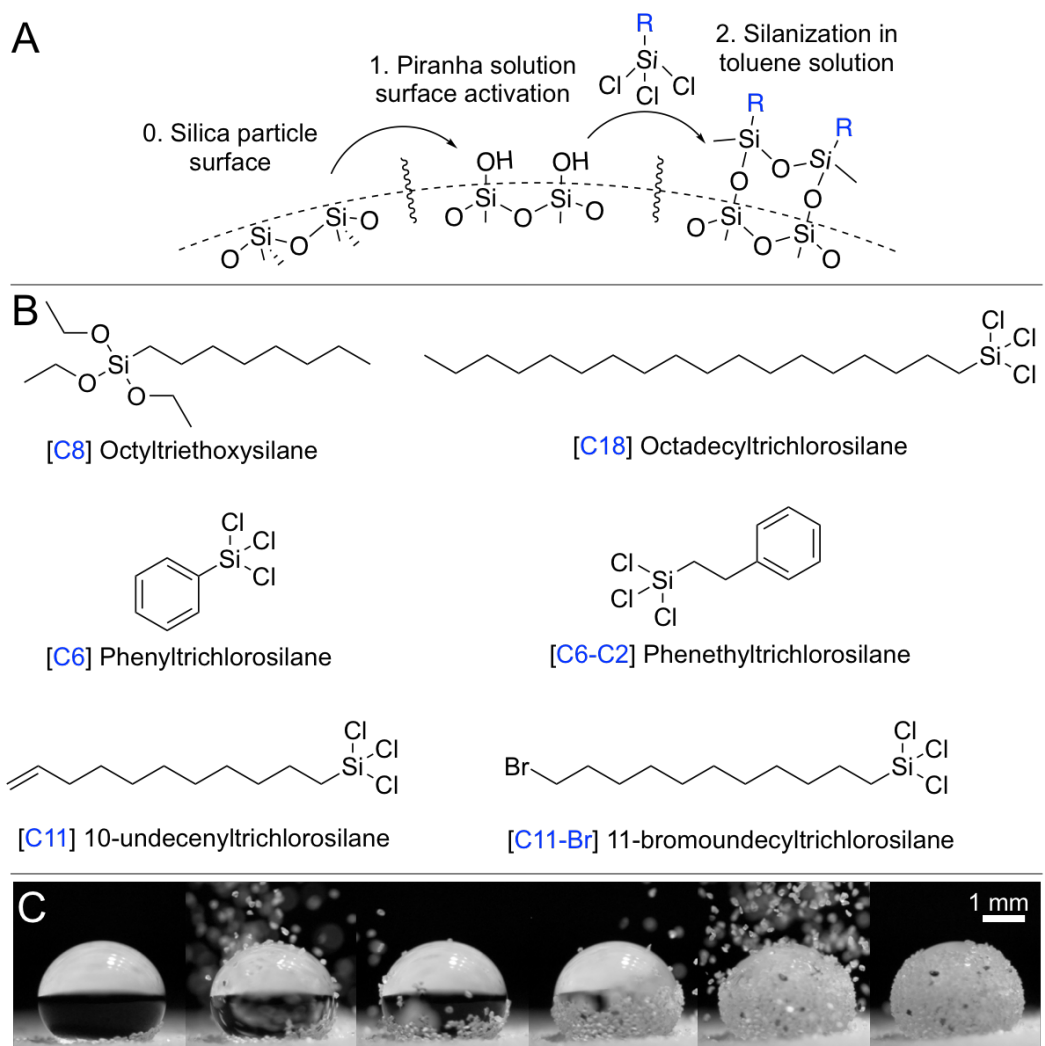

**Fig. 1. Fabricating a liquid marble.** (**A**) Silanization process. (**B**) Various hydrophobic surface compositions obtained using silanes. (**C**) Snapshots from Movie S1 (Supplementary Information), where the hydrophobic particles are poured onto a water droplet to create a liquid marble. The particles slide along the air–water interface and cover the liquid surface bottom–up, thus covering the droplet.

We studied $\sigma$ as a function of the particle size by fixing the particles' chemical make-up to octadecyltrichlorosilane, C18 (Table 1). A linear relationship was observed between the particle size, $d_p$, and the surface particle density, according to the relationship, $\sigma \propto A.\rho_p.d_p/A$, where $\rho_p$ is the bulk density of the particles, and $A$ is the droplet's surface area (Fig. S1A). In addition, $\sigma$ was measured as a function of the particles' chemical composition (Table 1) by fixing particle size to 57 μm. A reasonable correlation was observed between $\sigma$ and the receding contact angles (Fig. S1C) because during the rolling of liquid marbles, particles are slightly pushed into the liquid, and the lower receding angles prevent them from being expelled, thereby increasing $\sigma$. Next, the interparticle friction was characterized, which arises from their chemical composition, topography (Fig. 2), and loading (*35, 44*). The angles of repose, $\theta_{rep}$, of the particulate cones formed by dropping the particles from a funnel (Table 1) were measured, and the tangent of the highest values of those angles yielded the static friction coefficient, $\mu = \tan\theta_{\text{rep}}$. Further, particle compressibility, defined as the change in the volume of a mass of particles under applied stress, was characterized. Compressibility can also be correlated with the variance in the angle of repose measurements for a given particle type ($\Delta\theta_{\text{rep}}$). For example, fumed silica particles with fuzzy nanostructure (Figure 2C) and characteristic size of 7 nm presented a wide range of angles of repose ($\Delta\theta_{\text{rep}} = 41°$) depending on how tightly they were packed, whereas all other particles exhibited significantly lower compressibility. Hereafter, the particles are referred to with a simple code X|Y, where X refers to their chemical composition (e.g., C8 and C18) and Y refers to the particle size in microns (Table 1).

**Table 1. Material characterization of the functionalized silica particles used in this study**: characteristic dimension, apparent contact angles for 10-µL water droplets advanced and receded at 0.2 µL·s$^{-1}$, surface particle density, and angle of repose of cones formed with particles.

| Exp. | Code | Functional group | Particle size, $d_p$ (µm) | Adv. angle, $\theta_{\text{adv}}$ (°) | Rec. angle, $\theta_{\text{rec}}$ (°) | Surface particle density, $\sigma \pm$ std. error (kg/m²) | Angle of repose, $\theta_{\text{rep}} \pm \Delta\theta_{\text{rep}}/2$ (°) [friction] |
|---|---|---|---|---|---|---|---|
| | water | [water droplet] | 0 | 135 (H-glass) | 110 (H-glass) | – | – |
| Varying coating (57 µm) | C8\|57 | Octyltriethoxy | 57 | 81 | 50 | 0.042 ± 0.000 | 43.0 ± 3.0 |
| | C18\|57 | Octadecyl | 57 | 114 | 102 | 0.030 ± 0.003 | 40.3 ± 2.5 |
| | C6\|57 | Phenyl | 57 | 101 | 73 | 0.039 ± 0.004 | 47.3 ± 2.3 |
| | C6-C2\|57 | Phenethyl | 57 | 98 | 59 | 0.048 ± 0.004 | 42.7 ± 3.0 |
| | C11\|57 | 10-undecenyl | 57 | 99 | 50 | 0.056 ± 0.004 | 44.8 ± 2.5 |
| | C11-Br\|57 | 11-bromoundecyl | 57 | 107 | 58 | 0.052 ± 0.000 | 41.5 ± 2.5 |
| | water | [water droplet] | 0 | 135 (H-glass) | 110 (H-glass) | – | – |
| Varying particle size (C18) | C18\|0.007 | Octadecyl | 0.007 | 114 | 102 | 0.023 ± 0.000 | 76.5 ± 20.5 |
| | C18\|0.5 | Octadecyl | 0.5 | 114 | 102 | 0.064 ± 0.009 | 50.5 ± 7.6 |
| | C18\|3 | Octadecyl | 3 | 114 | 102 | 0.006 ± 0.000 | 44.1 ± 3.0 |
| | C18\|57 | Octadecyl | 57 | 114 | 102 | 0.030 ± 0.000 | 40.3 ± 2.5 |
| | C18\|90.5 | Octadecyl | 75–106 | 114 | 102 | 0.145 ± 0.006 | 42.1 ± 3.0 |
| | C18\|128 | Octadecyl | 106–150 | 114 | 102 | 0.179 ± 0.009 | 40.5 ± 3.0 |
| | C18\|181 | Octadecyl | 150–212 | 114 | 102 | 0.255 ± 0.006 | 40.3 ± 3.0 |
| | C18\|231 | Octadecyl | 212–250 | 114 | 102 | 0.233 ± 0.007 | 42.5 ± 3.0 |
| | C18\|275 | Octadecyl | 250–300 | 114 | 102 | 0.370 ± 0.019 | 44.2 ± 2.5 |
| | C18\|300 | Octadecyl | 300 | 114 | 102 | 0.309 ± 0.036 | 41.5 ± 3.5 |
| Super-hydro-phobic (SH) | SH\|128 | Octadecyl + Glaco™ layer | 128 | >150 | >150 | 0.076 ± 0.001 | 40.5 ± 3.0 |

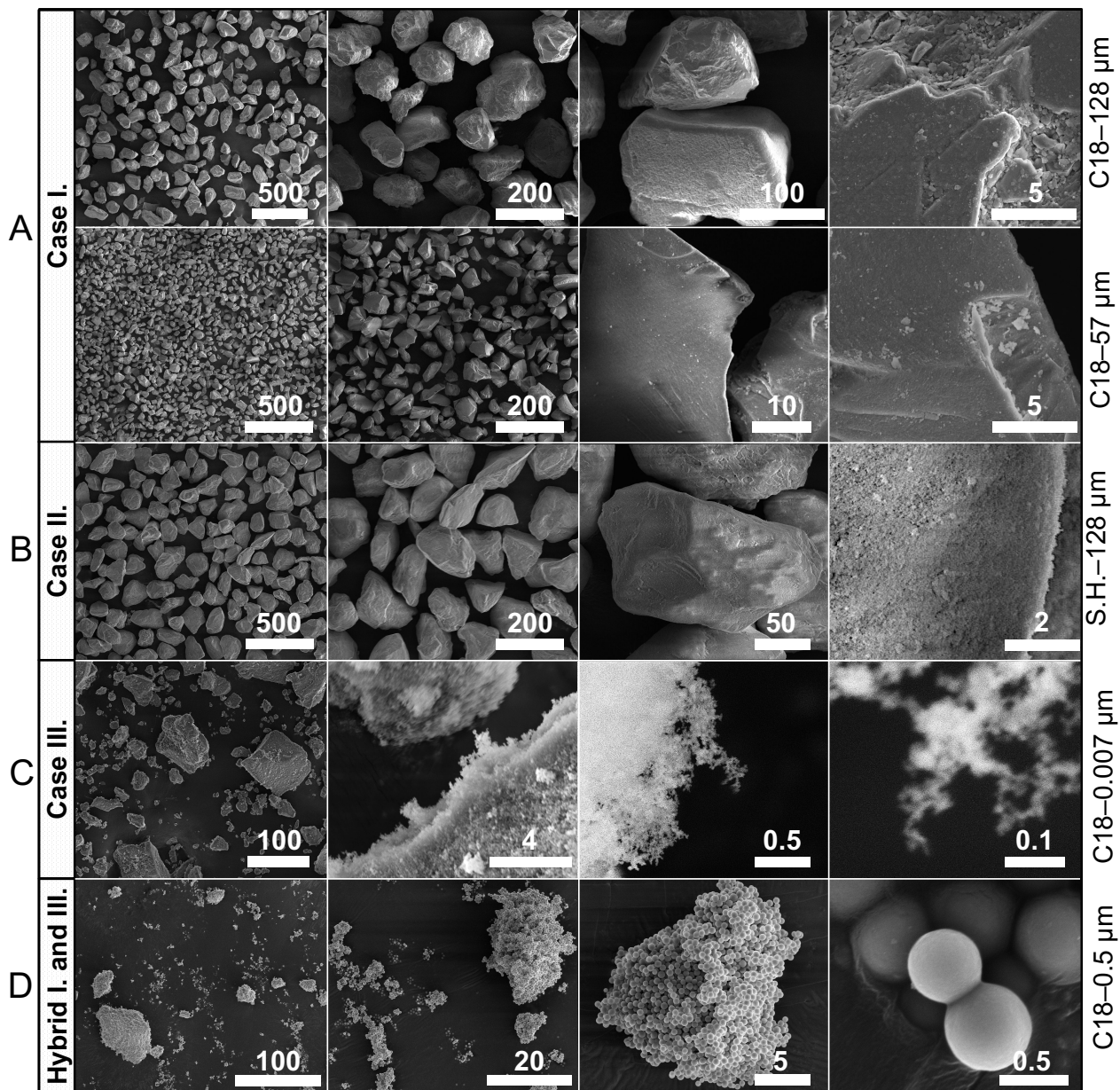

**Fig. 2. Representative scanning electron micrographs of a subset of particles used in this study.** Scale bars are in micrometers. The last column presents the nomenclature pertaining to the particles based on their chemical composition and size (Table 1). **(A)** Liquid marbles formed with these particles result in Case I on deflation; **(B)** liquid marbles formed with these superhydrophobic particles result in Case II; **(C)** liquid marbles formed with fuzzy nanoscale particles that tend to agglomerate result in Case III. Note: The cases are extensively discussed in the following sections, and representative electron micrographs for the remaining particles are presented in the Supplementary Information (Fig. S2).

*Fates of deflating liquid marbles*

Next, water loss was induced in the liquid marbles by allowing them to evaporate under laboratory conditions: temperature of $23 \pm 1°C$ and $60 \pm 2\%$ relative humidity. A precision mass-balance was used to monitor time-dependent changes in the water content, which was tracked as the liquid mass fraction, $m/m_o$, where $m$ and $m_o$ denote the instantaneous and initial masses of the liquid, respectively. This experimental setup was essentially equivalent to slowly removing water from a liquid marble with a capillary such that only the liquid was removed and not the particles. In addition to monitoring the mass, the concomitant structural changes in the stressed liquid marbles were observed via time-lapse imaging (Fig. 3). The curved particle-laden surface of a liquid marble was observed to experience a tangential compression as the liquid evaporates, which is analogous to the compression of a flat liquid–particle–vapor interface induced by a Langmuir–Blodgett trough reported previously (*43*). Interestingly, the gradual deflation of liquid marbles herein, which are formed from particles of sizes varying over four orders of magnitude and drastically different chemical compositions, revealed three general cases. These cases could be classified on the basis of liquid–particle (L–P) adhesion, interparticle (P–P) friction. These general cases are described below followed by examples.

- **Case I—constant surface area** (Fig. 3B): this case involves liquid marbles formed with particles with high liquid–particle adhesion (intermediate to low $\theta_{\text{rec}}$ values) and moderate to low interparticle friction ($\theta_{\text{rep}}$). As they lose the liquid, the marbles maintain the particulate monolayer and preserve their surface area, resulting in significant structural deviation from sphericity.
- **Case II—particle ejection** (Fig. 3C): this case involves liquid marbles formed with superhydrophobic particles having low liquid–particle adhesion (highest $\theta_{\text{rec}}$ values) and low interparticle friction ($\theta_{\text{rep}}$). As the liquid is lost, the marbles maintain high sphericity and particulate monolayer. To accomplish this, they eject particles from their surface.
- **Case III—thickening of the particle layer** (Fig. 3D): this case involves liquid marbles formed with particles having low liquid–particle adhesion (highest $\theta_{\text{rec}}$ values) and high interparticle friction and compressibility. As the liquid is lost, the liquid meniscus dewets particles, which remain adhered to their neighbors, thereby thickening the particle layer and increasing $\sigma$; marbles maintain high sphericity.

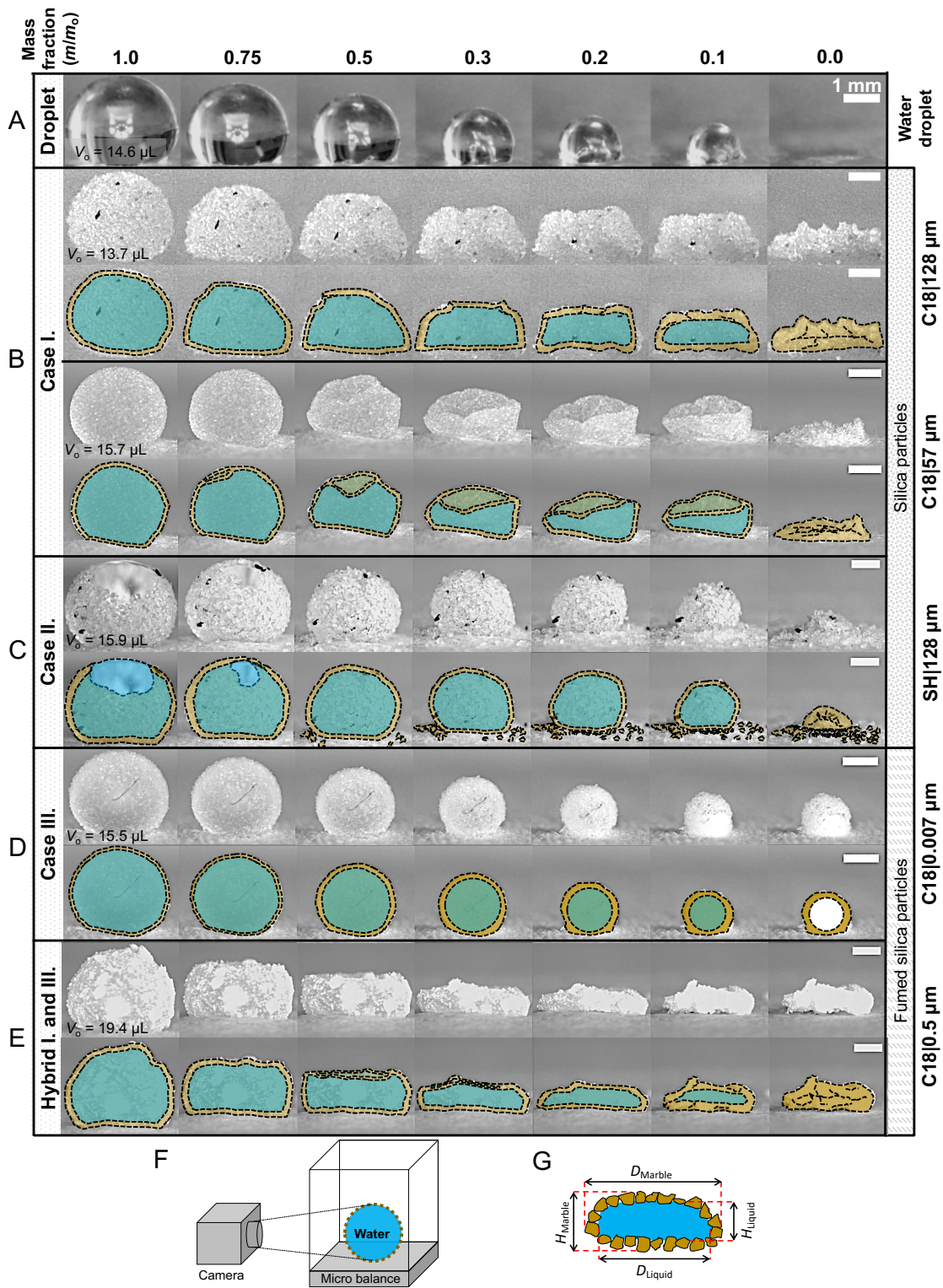

**Fig. 3. Representative snapshots and image analysis of evaporating liquid marbles formed with particles of sizes varying over four orders of magnitude (7 nm–300 μm) and drastically different chemical compositions. (A)** Control case: evaporating water droplet placed on a hydrophobic glass slide

(H-glass). **(B)** Case I: evaporating liquid marble maintains constant surface particle density and surface area, which requires its shape to deviate from that of a sphere. **(C)** Case II: evaporating liquid marble maintains high sphericity and constant surface particle density. In doing so, it ejects particles from its surface as the surface area decreases. **(D)** Case III: evaporating liquid marbles maintain sphericity as they shrink and the surface particle density increases owing to the compression and layering of particles at the liquid interface. After the liquid evaporates, a hollow shell is left behind. **(E)** Hybrid Cases I–III: the liquid marble behaves as a hybrid in Cases I and III and exhibits significant deviation from sphericity and thickening of the particulate layer. Schematics of the **(F)** experimental setup and of **(G)** a liquid marble depicting the particles and liquid within. (Note: Additional time-lapse data for the remaining liquid marbles are shown in Fig. S3 and Movie S2).

Time-lapse imaging experiments revealed that in **Case I**, particles exhibit high adhesion to the liquid, which keeps them stuck to the interface until the very final stages of evaporation ($m/m_o < 0.10$). By then, the liquid marble is so deformed that the meniscus recedes, possibly even detaching completely from small patches of wet particles. Another striking feature of Case I is the manner in which the marbles deform. For larger particles, the stiffness and weight of the particulate monolayer is so high that it results in the marbles' collapse, followed by the deflation of the hemispherical object (Fig. 3B top). For smaller particles ($\leq 57$ μm), the dome collapses first, and the marble subsequently wrinkles as it deflates (Fig. 3B bottom).

In **Case II**, the apparent contact angles at the particle–liquid–vapor interface are the highest (e.g., SH|128—Table 1 with $\theta_{adv} > 150°$ and $\theta_{rec} > 150°$). SH|128 exhibits superhydrophobicity because silica particles are functionalized with a C18 silane layer followed by multiple Glaco™ coats (Fig. 2B rightmost frame). Thus, the combination of the hydrophobic coating and nanoscale roughness resulting from the use of Glaco™ renders the particles superhydrophobic, yielding ultralow liquid–particle adhesion(45, 46). In addition, the interparticle friction is low, $\theta_{rep} = 40.5°$. As the liquid volume decreases, the particles get compressed against each other, generating an ejection force that expels some particles from the marble surface (discussed in detail in the next section). Consequently, the sphericity of the marble does not significantly deviate from its initial shape during evaporation (Fig. 3C).

**Case III** differs from the previous two cases owing to the high interparticle friction and compressibility, i.e., their ability to form dense particle layers under compression (Fig. 2C). Indeed, high compressibility is evidenced by the high variance in the angle of repose of the particles as a function of particle compression ($\theta_{rep} = 76.5°$ and $\Delta\theta_{rep} = 41°$ – Table 1). As liquid marbles formed with these particles evaporate, particles are pushed against each other and a fraction get dewet. Due to the high interparticle friction, these particles are not ejected and they result in multilayering, increasing the packing density. This can be easily observed in the 7-nm fumed silica particles in Fig. 3D wherein the small black cotton fiber on the marble provides a contrasting object that becomes gradually buried as the layer thickens. Although coated with the same hydrophobic molecule (C18: $\theta_{adv} = 114°, \theta_{rec} = 102°$) as the other particle sizes, these particles display superhydrophobicity owing to their nanoscale roughness, which is characterized by the contact angles $\theta_{adv} > 150°$ and low hysteresis (47, 48). This occurs because they are highly branched, which traps air between the particles (49) (Fig. 2C). During this process, liquid marbles maintain highly spherical shapes.

*Analytic framework*

Here, the analytic framework is described for predicting the potential final states of liquid marbles (Cases I–III) that are based on liquid–particle (L–P) and particle–particle (P–P) interactions. First, note that liquid marbles are typically formed at length scales below the capillary length, $\lambda_C = \sqrt{\gamma_{LV}/(\rho_L \times g)}$, where $\gamma_{LV}$ denotes the surface tension of the liquid–air interface, $\rho_L$ denotes the density of the liquid, and $g$ denotes the acceleration due to gravity (for water, $\lambda_C \approx 2.7$ mm) (*48*). For length-scales $\leq \lambda_C$, the contributions of weight and buoyancy are negligible compared to those of capillary forces. Thus, a freshly made liquid marble assumes a spherical shape under the influence of tensions at the liquid–vapor and liquid–solid interfaces (Fig. 4). Owing to evaporation, the liquid interface in contact with the particles tends to recede, such that the apparent contact angles at the solid–liquid–vapor (S–L–V) interface transition from $\theta_{adv} \to \theta_{rec}$. The receding meniscus pulls onto the particles with its surface tension, thus applying an interfacial force, $F_{int}$. This force compresses the particles against one another, generating the compression force, $F_c$. Owing to the particles' finite size, irregular packing and surface topography, this compressive force results in an ejection force, $F_{ej}$, which tends to expel the particle away from the liquid surface, and an interparticle friction force, $F_{fr}$, which counteracts the ejection force.

For simplicity, we considered a force balance along the tangent at P–P contacts (Fig. 4B). The magnitude of the compression force, $F_c$, relates to the interfacial force as, $F_c = -F_{int} = -\gamma_{LV} \times 2\pi r_{wet} \times \cos 0°$, where $r_{wet}$ is the radius at the particle's wetting perimeter, which is maximum when $r_{wet} = r_P$ (coincident with particle radius, at the equator), and the angle is 0° because the force is locally tangential to the liquid–air interface. $F_c$ can be projected on the P–P contact axis, giving rise to the **ejection force**, $F_{ej} = F_c \times \sin\alpha$, or:

$$F_{ej} = -\gamma_{LV} \times 2\pi r_{wet} \times \sin\alpha \qquad (1)$$

where $\alpha$ is the particle relative position, given by the angle between the particle being ejected and its neighboring particles (Fig. 4B). For perfectly aligned horizontal particles, $\alpha = 0°$, which gives no ejection force, whereas, the higher the particle misalignment, the higher the ejection component of the compression force.

The **adhesion force**, $F_{adh}$, or L–P adhesion force is the projection of the $F_{int}$ onto the P–P contact axis:

$$F_{adh} = \gamma_{LV} \times 2\pi r_{wet} \times \cos\theta_{adh} \qquad (2)$$

Conversely, the $F_{int}$ in this case is tangential to the local liquid meniscus at the S–L–V contact point (Fig. 4C). This angle relates to the wettability of the particle, and can be approximated by the receding angle, $\theta_{rec}$. However, the liquid meniscus can get pinned at reentrant or doubly reentrant features (*50, 51*) on the particle surface, which can in reality decrease even more the meniscus angle relative to the P–P contact axis, $\theta_{adh}$, thus maximizing $F_{adh}$.

The **friction force**, $F_{fr}$, is given by the normal component of the compression force projected onto the P–P contact axis, $F_N$, multiplied by the coefficient of static friction between particles, $\mu$. And, $\mu$ is equal to the tangent of the angle of repose, $\mu = \tan\theta_{rep}$, where $\theta_{rep}$ is the experimentally measured angle of repose of the particles (Table 1, Methods) (*35*). Thus, $F_{fr} = \mu \times F_N$, which on substituting $F_N = F_c \times \cos\alpha$ yields,

$$F_{fr} = (\tan\theta_{rep}) \times (\gamma_{LV} \times 2\pi\, r_{wet} \times \cos\alpha) \qquad (3)$$

Together, these forces dictate the fate of the particles (Fig. 5) and the potential final states of the liquid marbles (Fig. 4D-F). We introduce the **total normalized force**, $F_T$, as the sum of these forces (Eqs. 1-3) normalized by the liquid wet perimeter, $P_{wet} = 2\pi\, r_{wet}$, such that $F_T = \frac{1}{P_{wet}}(F_{ej} + F_{adh} + F_{fr})$, which after some rearrangement yields,

$$F_T = \gamma_{LV}\,(\cos\theta_{adh} + \cos\alpha \times \tan\theta_{rep} - \sin\alpha) \qquad (4)$$

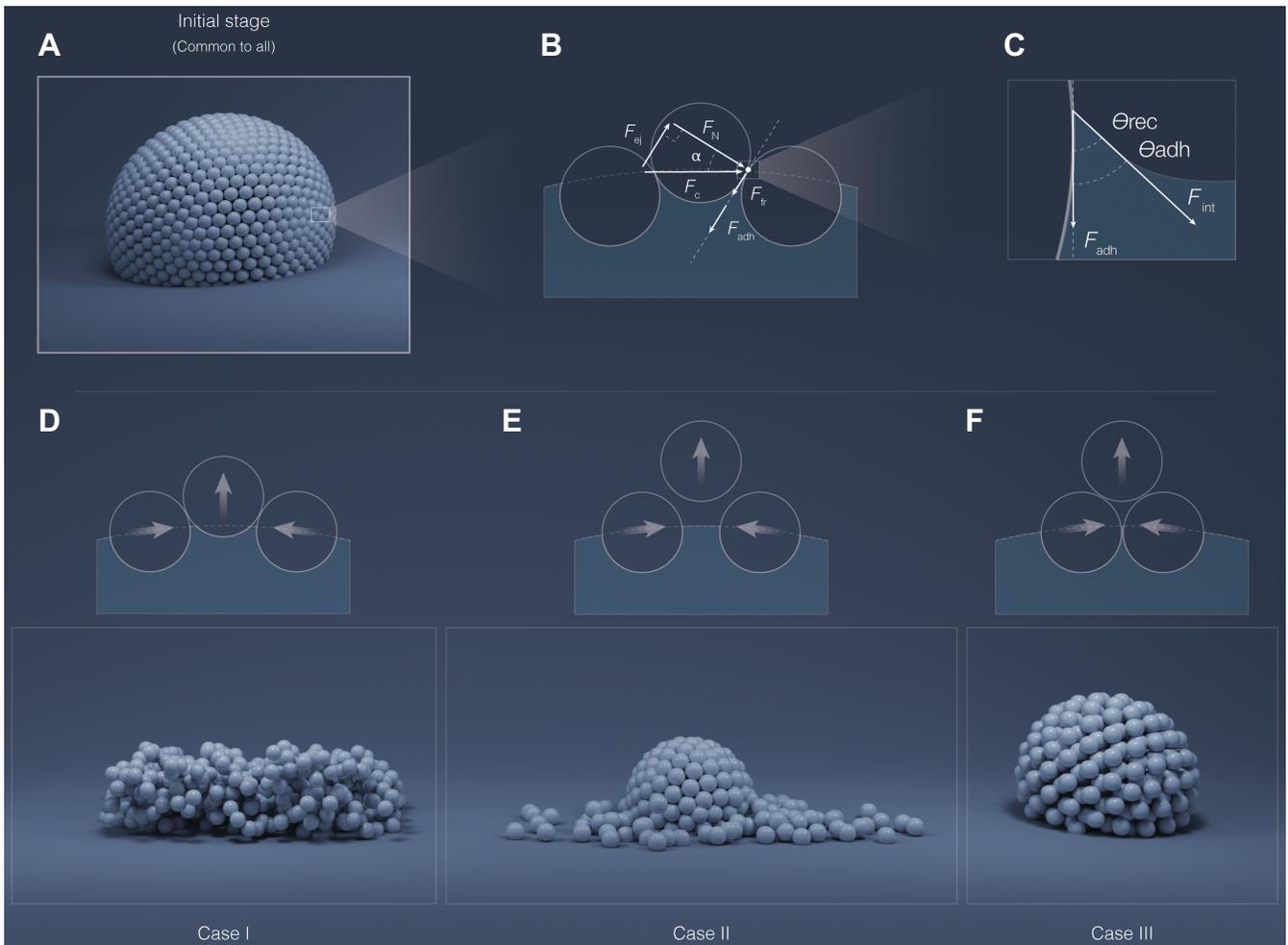

**Fig. 4. Analytic framework: interfacial force balance for a simplified system of spherical particles placed on an evaporating liquid marble.** **(A)** Illustration of the initial state of a liquid marble. **(B)** Force balance for the center particle applied at the contact point between two particles. As the water evaporates, the receding liquid meniscus pulls onto the particles owing to water's surface tension, generating a tangential force, $F_{int}$, which compresses the particles, $F_c$, such that $F_c = -F_{int}$. This compression force results in the ejection force, $F_{ej}$. The ejection force is countered by the forces of liquid–particle (L–P)

adhesion, $F_{adh}$, and interparticle (P–P) friction, $F_{fr}$. **(C)** Shows the actual microscopic local curvature of the liquid meniscus at the particle interface, which gives rise to $F_{adh}$. The possible outcomes for the center particle based on the force balance are: **(D)** the particle stays pinned to the liquid, characterizing Case I, **(E)** the particle gets ejected, Case II, and **(F)** the particle gets unpinned from the liquid but is held by the neighboring particles due to high P–P friction, Case III. Illustration created by Ivan Gromicho, Scientific Illustrator, Research Communication and Publication Services, Office of the Vice President for Research, King Abdullah University of Science and Technology.

When $F_T < 0$ the total normalized force points away from the liquid. In this case, as the meniscus starts to recede, the adhesion force plummets to zero, and $F_{ej}$ launches the particle off, characterizing Case II. Conversely, if $F_T > 0$, the particles are not ejected, and we can have Case I, which is characterized by high liquid-particle (L–P) adhesion, or Case III where the interparticle (P–P) friction dominates. Fig. 5 shows a compilation of possible scenarios for different L–P and P–P properties, where the colored areas indicate the regions of properties for each of the Cases I–III. For instance, as the meniscus relative angle, $\theta_{adh}$, increases, meaning that effectively the hydrophobicity increases, the total normalized force goes from positive to negative, or from a region where Case I prevails to a region of Case II (Fig. 5A). However, if P–P friction increases, the scenario shifts towards the Case III region. Case II is characterized by the highest hydrophobicity ($\theta_{rec} > 150°$, note that we assume $\theta_{adh} \approx \theta_{rec}$), but moderate to low P–P friction ($\theta_{rep} = 40.5° \pm 3.0°$), conferring to its particles extreme low L–P adhesion, and low enough P–P friction that the particles can easily detach from the liquid and be ejected outwards without being stuck to neighboring particles. The next important factor into consideration in our force balance, is the particle position relative to its neighbors, $\alpha$. The more aligned the particles are, the lower the magnitude of the ejection force, $F_{ej}$. Consequently, the higher the contribution of L–P adhesion and P–P friction to a positive (towards liquid) resultant force (Figs. 5B, C, D, F).

Interestingly, hybrids of Cases I & III are also possible, e.g., C18|0.5 lying in the overlapping regions (Fig. 5A, D, E, F). This hybrid case has a $\theta_{rep} = 50.5° \pm 7.6°$ (Table 1), effectively not as high as Case III marbles (e.g., C18|0.007 with $\theta_{rep} = 76.5° \pm 20.5°$), but higher than Case I marbles (all coatings with $40° < \theta_{rep} < 47°$). The high P–P friction of C18|0.007 and C18|0.5 is caused by their nanoscale roughness, which also contributes to their hydrophobicity due to air entrapment.

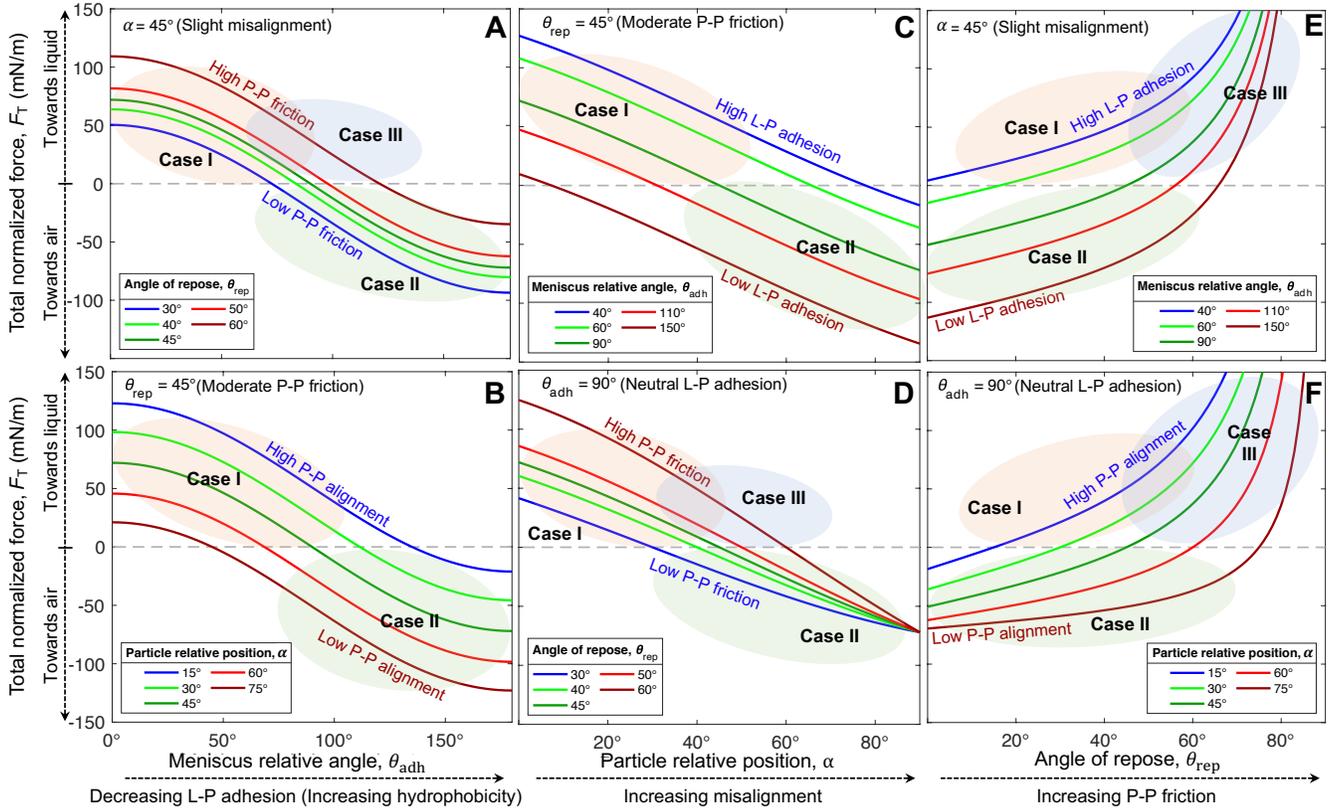

**Fig. 5. Analytic framework predictions.** Resultant force normalized by the particle wet perimeter, $F_T$, as function of (**A-B**) meniscus relative angle, $\theta_{\text{adh}}$, which relates to the L-P adhesion, (**C-D**) particle relative position, given by the angle $\alpha$, and (**E-F**) angle of repose of particles, $\theta_{\text{rep}}$, which relates to the P–P friction. The colored areas indicate the regions of properties for each of the Cases I–III. The boundaries of regions have not been clearly defined by our experiments. Although, the overlapping region between Case I & III was experimentally observed (C18|0.5). Notice that generally, shifting the properties the positive (Y-axis) left side in each plot tend to Case I, while case III is represented by general shifts towards the positive (Y-axis) right side. Whereas, Case II is represented by the negative values of the resultant force.

*Modeling of liquid marble evaporation*

In this section and the next we apply the insights from Cases I–III of the present study's analytic framework to capture the potential final states of evaporating liquid marbles using a model that harnesses a single geometric parameter, $k_e$.

In the simplest control case, the evaporation rate of a water droplet evaporating under normal temperature and pressure (NTP) conditions is limited by the diffusion of vapor in the surrounding air (*8, 52, 53*). Analogously, this is also the limiting step for Cases I–III until a certain moment during the evaporation (details in the Discussion section). Following those reports, the local radial evaporation flux from the liquid marbles can be expressed as $j = j_o/R$, where $R$ is the droplet/marble radius, and $j_o$ depends on the diffusion coefficient of water in air ($\mathfrak{D}_w$) as well as the saturated and ambient vapor concentrations, $\rho_{\text{sat}}$, and at a long distance from the droplet, $\rho_\infty$, over the liquid density of water, $\rho_L$:

$$j_o = \mathfrak{D}_w \frac{\rho_{\text{sat}} - \rho_\infty}{\rho_L} \tag{5}$$

Thus, the evaporation rate can be expressed as the product of the radial flux and the droplet or marble's exposed surface area, $A$.

$$\frac{dV}{dt} = -\mathfrak{D}_w \frac{\rho_{\text{sat}} - \rho_\infty}{\rho_L} \frac{A}{R} \tag{6}$$

Next, we write the equation in terms of liquid mass fraction, $m/m_o$, where $m$ represents the mass at a given time, and $m_o$ represents the initial mass of the liquid.

$$\frac{dV}{dt} = \frac{m_o}{\rho_L} \frac{d(m/m_o)}{dt} \tag{7}$$

Considering the structural deformation of the marble as the water is removed, e.g., in Case I, $A/R$ is expressed as a function of a characteristic diameter, $D_{\text{char}}$. Approximating the liquid marble to be a virtual hemisphere, where $A$ is area of the hemisphere exposed to air, we obtain

$$\frac{A}{R} = \frac{2\pi R^2}{R} = \pi D_{\text{char}}. \tag{8}$$

Thus, the characteristic diameter is representative of the apparent size of the liquid front in the marble, i.e., not considering the particles' contribution to the volume. Next, consider the concept of sphericity, $\varphi_{\text{object}}$, which describes the dissimilarity in the area of an object from that of a sphere of the same mass and density as $\varphi_{\text{object}} = A_{\text{sphere}}/A_{\text{real}}$. Analogously, a dissimilarity factor, $\psi$, is introduced to describe the deviation in the characteristic diameter of the liquid front inside the marble from that of the hemispherical droplet as $\psi = D_{\text{hemisphere}}/D_{\text{char}}$, where $D_{\text{hemisphere}} = [12\, m_o(m/m_o)/(\rho_L \pi)]^{1/3}$. During the lifetime of a liquid marble, $\psi$ represents the deflation and distortion. Based on our experimental data on the evaporation of liquid marbles introduced in Fig. 3 and presented in Fig. 6B, D, we considered that $\psi$ may adopt the following functional form, $\psi = (m/m_o)^{k_e}$, where $k_e$ is a geometric parameter related to the evolution of the shape of the marble and the particulate layer. For example, for the control case of a spherical droplet that evaporates with a constant contact angle, $k_e = 0$, implying that the shape of the exposed interface is not a function of the mass fraction, i.e., constant sphericity. Applying this logic to liquid marbles, the characteristic diameter at any stage of evaporation is given by

$$D_{\text{char}} = \left(\frac{12\, m_o}{\rho_L \pi}\right)^{\frac{1}{3}} \left(\frac{m}{m_o}\right)^{\frac{1-3k_e}{3}}. \tag{9}$$

Substituting Eqs. 7–9 in Eq. 6, the final expression for the evaporation of liquid marbles in terms of liquid mass fraction, ambient properties, and the geometric parameter is obtained as

$$-\frac{d(m/m_\text{o})}{dt} = \frac{\mathfrak{D}_w(\rho_\text{sat}-\rho_\infty)}{m_\text{o}^{\frac{2}{3}}}\left(\frac{12\pi^2}{\rho_L}\right)^{\frac{1}{3}}\left(\frac{m}{m_\text{o}}\right)^{\frac{1-3k_\text{e}}{3}}. \tag{10}$$

*Experimental results and model fitting for the evaporation of liquid marbles*

As described in Figs. 1–3 and Table 1, we formed liquid marbles with 10 μL of water using functionalized silica particles in the size range of 7 nm–300 μm; the particles' receding contact angles for water varied in the range of 50°–150°, and P–P interactions were characterized by repose angles in the range of 40°–76°. During their evaporation, some liquid marbles exhibited higher evaporation rates compared to the controls (10-μL water droplets), whereas at other instances, the marbles exhibited lower evaporation rates (Fig. 6B, D). By fixing the particle size to 57 μm and varying their chemical composition, we observed higher evaporation rates compared to those of the controls (Fig. 6A–B). For Case I, the wiggling behaviors corresponded to actual variations in the evaporation rates owing to particle compression and adjustment together with the marbles' structural deformations caused by water loss. Clearly, this effect was absent in the case of water droplets.

Next, the effects of particle size on evaporation rates were studied by fixing the chemical composition to C-18. While there was no clear dependence on particle size, liquid marbles formed with 7-nm fumed silica particles consistently exhibited slower evaporation than others (Fig. 6C). Interestingly, for these marbles, the evaporation rates decreased to even below that of the control cases (water droplets) at $m/m_\text{o} < 0.3$ (Fig. 6D). The experiments revealed that this was caused by the transition of water evaporation from an air-limited vapor diffusion to a particulate-layer-limited vapor diffusion. This resulted from the thickening of the particle layer, which constricted vapor transport. Moreover, this feature underlies a decrease in the evaporation rates of 7- and 500-nm-sized marbles in contrast to other marbles in the range of $m/m_\text{o} \approx 0.75 - 0.65$. Finally, the evaporation rates for the liquid marbles formed with superhydrophobic particles (SH|128 μm, in pink) were punctuated with significant oscillations owing to particle ejection.

Our model, equipped with a single geometric parameter, $k_\text{e}$, accurately described all potential final states of the liquid marbles during evaporation, i.e., Cases I–III, along with their hybrids (Figs. 6E–H). The predicted $m/m_\text{o}$ values ranged within a coefficient of determination of $R^2 > 0.98$ with respect to the experimental data (Figs. 6E, G). In its ability to capture the evaporation rate as a function of $m/m_\text{o}$, the model satisfactorily correlated with the experimental data, yielding $R^2 > 0.85$ (Figs. 6F, H) in most cases. Finally, the model adequately captured the average behavior of liquid marble evaporation in Case II (for SH|128 μm). The poor correlation ($R^2 = 0.28$) was caused by the high instantaneous variance in the evaporation rate owing to particle ejection, as explained above.

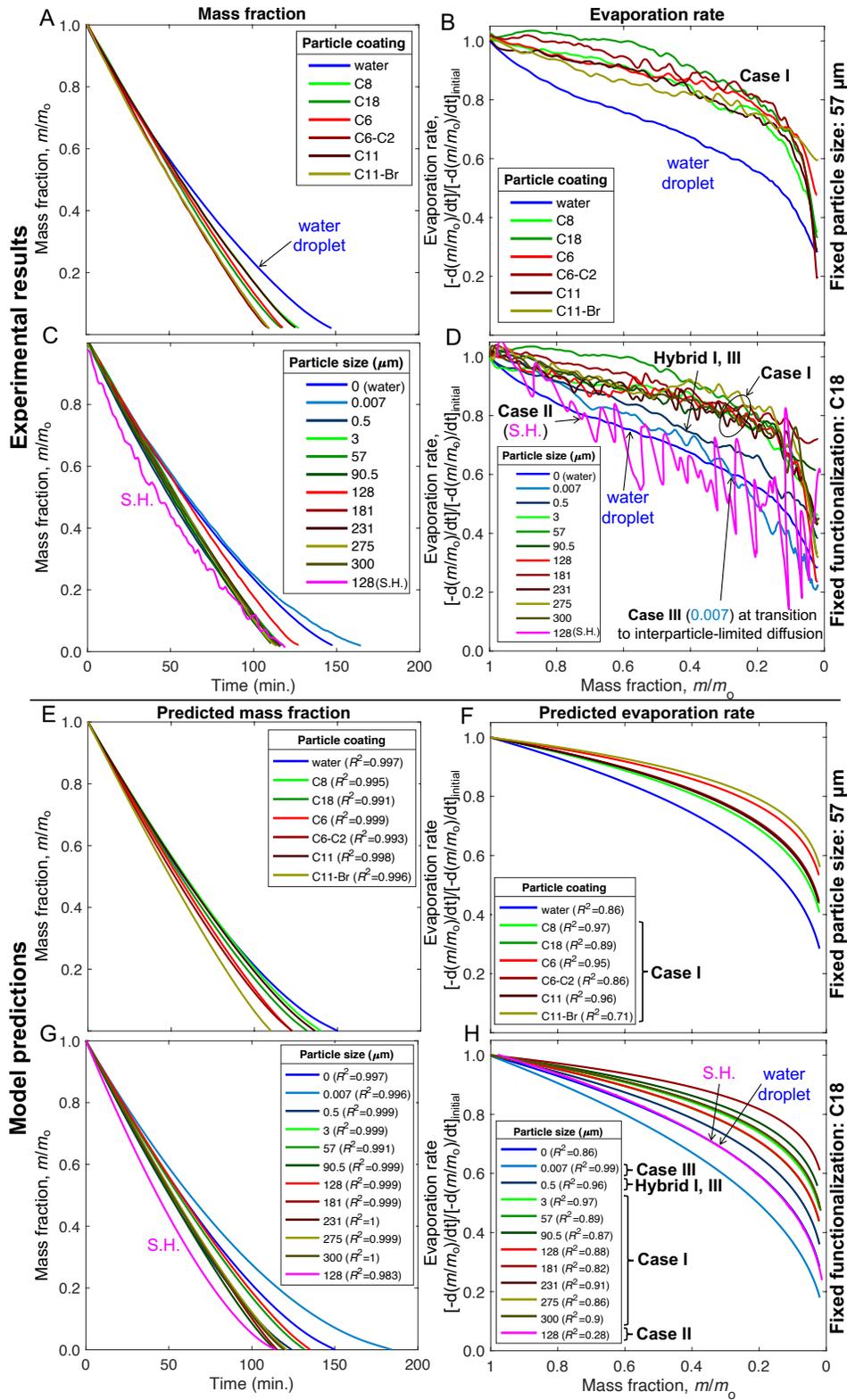

**Fig. 6. (A–D)** Experimental results and **(E–H)** model predictions for the liquid mass fraction and evaporation rates of liquid marbles formed with 10-μL water and evaporated at 23±1°C and 60 ± 2% RH. **(A–B, E–F)** Effects of varying the chemical make-up of 57- μm silica particles. **(C-D, G-H)**

Effects of varying the particle size from 7 nm to 300 µm while keeping the chemical make-up fixed to the C18 coating (except for the superhydrophobic case that received an additional coating of Glaco™). The legends "water" and "0" refer to sessile water droplets placed on the smooth hydrophobic glass surfaces (H-glass). The mass fraction data was truncated at $m/m_o = 0.02$ owing to the disintegration of liquid marbles and ensuing complications (Fig. S6).

The marbles in **Case I** exhibited a fitted geometric parameter of $k_e = 0.16$ with a standard deviation of 0.04 (Fig. 7A), which is independent of the apparent contact angles pertaining to the particle's chemical composition (Fig. 7B–C). Thus, $D_{\text{char}} > D_{\text{hemisphere}}$ and the evaporation rates were higher than that of water droplets, which is reasonable owing to the significant structural deformation of the liquid marbles during evaporation resulting from high L–P adhesion.

**Case II** has $k_e \approx 0$ because of the minimal structural deformation of liquid marbles when the compression in the particulate layer is released via particle ejection; low P–P friction preempts multilayering. Thus, in this case, $D_{\text{char}} \approx D_{\text{hemisphere}}$.

For **Case III**, $k_e$ is negative; thus, the effective evaporation rate is lower than that of a bare hemispherical droplet, i.e., $D_{char} < D_{\text{hemisphere}}$. This is reasonable considering the limiting effect of the diffusion barrier created by multilayering because of high P–P friction.

Through this modeling approach, we were also able to pinpoint the behaviors of hybrid liquid marbles that simultaneously exhibited features of Cases I and III. For the 500-nm fumed silica particles, the $k_e$ value was intermediate relative to those of Cases I and III (Fig. 7A). These liquid marbles underwent structural deformation and experienced thickening of the particulate layer; since $k_e > 0$, the contribution of the former dominated the latter.

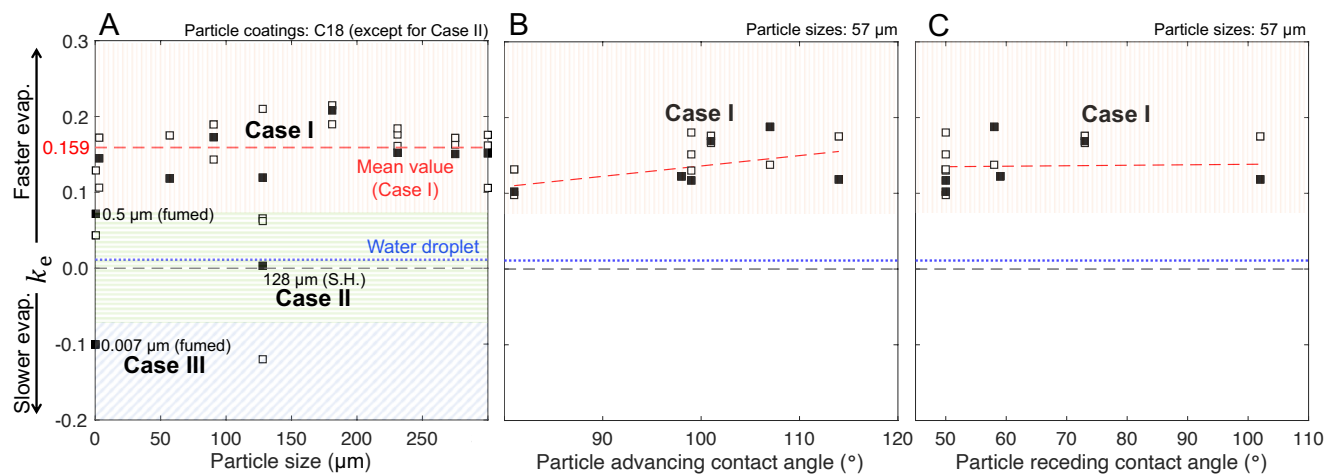

**Fig. 7. Correlations between the geometric parameter, $k_e$, and key particle characteristics.** Shaded regions represent Cases I–III. **(A)** Effects of particle size, maintaining the chemical composition as fixed at C18 with the exception of SH|128 that had the Glaco™ coating on top of a C18 layer. **(B)** Advancing and **(C)** receding intrinsic contact angles for 57-µm-sized particles with coatings ranging from hydrophilic to hydrophobic. Black-filled squares are points obtained with the mean values of evaporation runs,

whereas white-filled squares represent replicates. Note: In these plots, $k_e = 0$ represents the perfectly hemispherical case, which the control case (water droplets, dotted blue line) approaches. The 0.5-μm (fumed) particles are the hybrid/intermediate between Cases I and III, wherein the deformation behavior is more pronounced than the layering behavior, thus exhibiting positive $k_e$.

**Discussion**

This study developed an analytic framework based on L–P and P–P interactions to predict the potential final states of liquid marbles when the liquid is gradually removed. Based on these insights, a single parameter model is developed that accurately captures the evaporation of liquid marbles formed with particles of sizes varying from 7 nm to 300 μm and chemical compositions ranging from hydrophilic to superhydrophobic. This section discusses some peculiarities of the observations and discusses associations with existing literature on this subject.

The most commonly reported liquid marbles in the literature are formed with water and belong to Case I because common hydrophobic particles exhibit intermediate $\theta_{rec}$ and intermediate P–P friction (Table 1) (*8*). These particles do not have hierarchical, reentrant, and/or nanoscale features, which prevents them from robustly entrapping air at the L–P contact or exhibiting particle-level superhydrophobicity (*54*). It should be noted that we are discussing the particle-level wetting behavior; at the macroscale, a flat surface with a layer of these particles will, obviously, exhibit superhydrophobicity owing to the entrapment of air between the particles. Thus, in the liquid marble configuration, the L–P interface for these particles would be in the Wenzel state (*55*), and the L–P adhesion would dominate over the ejection force (Case I). Then, these liquid marbles maintain constant surface area and monolayer during evaporation by undergoing significant structural deviation from sphericity. Thus, they exhibit faster evaporation rates than that of bare droplets along with $k_e > 0$ (Figs. 6B, D, 7A–C).

In contrast, if nanoscale surface roughness is realized onto the particle surface, e.g., if SH|128 particles are coated with Glaco™, they exhibit superhydrophobicity even at the particle level, which is characterized by $\theta_{adv} > 150°$ and $\theta_{rec} > 150°$ (Table 1 and Fig. 2C). Thus, in the liquid marble configuration, the L–P interface for these particles would entrap air between nanoscale asperities (Cassie states) (*56*). Interestingly, we were unable to find reports on liquid marbles formed with superhydrophobic particles, presumably because their extreme water repellence complicates their preparation, and even faint mechanical vibrations can damage them owing to the ejection of patches of particles. For example, a liquid marble formed with SH|128 particles lost a patch of particles as excess particles were swept off the evaporation plate (Fig. 3C). Then, the bare patch gradually filled up through particle rearrangement owing to evaporation, the compression force built up with further evaporation, and the particle ejection began. Notably, particle ejection ceased in the final stages of evaporation. This can be explained based on the inhomogeneities among the particles—in terms of structural and surface coating—such that the particles with higher water-repellence were ejected first, leaving behind less hydrophobic particles. Particle ejection due to low L–P and P–P friction ensures that the liquid marbles maintain a monolayer of particles and remain spherical (Case II). Thus, the evaporation rates are similar to those of the control cases (water droplets) and $k_e \approx 0$ (Figs. 6H, 7A). Furthermore, ejected particles landed on the mass balance, and their impacts caused instabilities in the recorded data (Figs. 6 C–D). Some fluctuations in the recorded data are caused by particle rearrangement.

If superhydrophobic particles have high P–P friction, the outcome is markedly different. Fumed silica particles (C18|0.007) with a dendritic structure formed larger, fluffier particles owing to their high P–P friction (Fig. 2C and Table 1)(*57*). Thus, in the liquid marble configuration, the hairy L–P interface entrapped air (*58*). When the liquid is removed, some particles dewet but remain connected to neighboring particles causing multilayering (*40, 59*). Interestingly, interparticle friction was so robust for these particles that, in one of the replicates, the liquid completely detached from the particle structure at $m/m_o < 0.1$, leaving a free-standing particulate dome (Fig. 3D). Remarkably, the increase in particle layer thickness was so pronounced due to multilayering that the interparticle vapor diffusion became the limiting step for evaporation (*8*). This was experimentally evidenced at $m/m_o \approx 0.3$ when the evaporation rate of the C18|0.007 liquid marble dropped below that of bare water droplets (Fig. 6D). Although our model did not directly predict this bottleneck transition—from vapor diffusion in air to interparticle vapor diffusion—it yielded negative values of $k_e$, signifying multilayering (Fig. 7A) and the reduction in the evaporation rate (Fig. 6H).

To summarize, L–P adhesion dominates over other forces in Case I; ejection force dominates over L–P and P–P friction in Case II, and P–P friction acts dominant in Case III (Table 2). Interestingly, for the hybrid Case I–III for the C18|0.5 marbles, a combination of structural deformation (Case I) and multilayering (Case III) (Fig. 3E) can be clearly observed and attributed to mild L–P adhesion and higher (but not highest) P–P friction (Fig. 5). In this case, particles are not ejected as in Case II owing to their branched structure, yielding moderate to high P–P friction as evidenced by their $\theta_{\text{rep}} = 50.5° \pm 7.6°$ (Table 1). Consequently, the $k_e$ value is positive (Fig. 7A) but lower in magnitude compared to typical Case I values, indicating a more pronounced structural deformation behavior than multilayering. Accordingly, the evaporation rates for Hybrids I–III are higher than that of the water droplet but not as high as that in Case I (Figs. 6D, H).

**Table 2. Case summary and outcomes of evaporation**

| Case | Geometric parameter, $k_e$ | L–P adhesion [$\theta_{\text{rec}}$]* | P–P friction [$\theta_{\text{rep}}$] | Dominant force | Particle outcome | Evaporation outcome vs. droplet |
|---|---|---|---|---|---|---|
| **I** | >0 | High [<150°]* | Moderate–Low [<~45°] | L–P adhesion | No layering | Faster |
| **II** | ~0 | Low [>150°]* | Low [<~40°] | Ejection force | Ejection | Same |
| **III** | <0 | Low [>150°]* | High [>~50°] | P–P friction | Layering | Slower |
| **Droplet** | ~0 | – | – | – | – | Same |

(*) – The threshold for the receding angle that separates Case I from Cases II–III has not been determined in this work; our experimental data suggests that it is at least $\theta_{\text{rec}} > 102°$. Thus, considering the particle surface inhomogeneities, we assume only the superhydrophobic cases, where $\theta_{\text{rec}} > 150°$.

Next, we examine some of the reported data in the literature through the framework of L–P and P–P interactions developed herein. For example, Bhosale et al. (*60*) have measured the evaporation rates of liquid marbles formed with a microscale polytetrafluoroethylene powder (μPTFE) and fumed silica nanoparticles coated with hexamethyldisilazane (nHMDS) and dimethyldichlorosilane (nDMDCS). They observed that the former evaporates faster, and the latter behaves similar to water droplets. μPTFE results can be classified as Case I based on the significant structural deformation of the liquid marble during evaporation, whereas both nHMDS and nDMDCS can be classified as Case III, wherein the larger surface area of nHMDS than nDMDCS caused severe reduction in vapor diffusion through the particle layer. Rouyer et al. (*8*) reported higher evaporation rates of liquid marbles formed with polystyrene particles (140 μm) silanized with perfluorodecyltrichlorosilane (FDTCS) compared to bare water droplets. Additionally, they reported significant structural deformation of marbles during evaporation, evidencing Case I. Erbil et al. (*33*) investigated the evaporation of liquid marbles prepared with polytetrafluoroethylene (PTFE) powder (5–9 μm) as a function of relative humidity. They reported structural deformation but with an overall lower evaporation rate compared to that of bare water droplets, indicating a hybrid of Cases I and III, in which the increased resistance to vapor diffusion results from the particle layer's thickening. Notably, they also reported that an increase in relative humidity (RH) resulted in higher resistance to evaporation compared to that of water droplets. This was attributed to the fact that PTFE particles tended to aggregate more in high humidity conditions, which could be due to the reduction in interfacial electrification (*36, 38*). Markedly, Case II is the least common in the literature; to the best of our knowledge, it has only been reported by McHale et al. (*42*) using spherical silica particles (75–1180 μm in size) coated with trimethylsilylchloride. Those particles were expected to have very low P–P friction, due to their sphericity, which facilitated their ejection from the liquid marble's surface. Experiments with different particle shapes, sizes, and surface chemistry and liquids are needed to extend this analytic framework for identifying precisely the boundaries between the cases and hybrids.

Lastly, the apparent volumes of the liquid marbles obtained from our image analysis and predicted values of the characteristic diameter, $D_{\text{char}}$, (Figs. S7B, D) are compared to underscore our model's ability of describing the liquid marble system. The obtained results demonstrate a reasonable correlation with $R^2 > 0.85$ for most marbles, except for the 57-μm-sized marbles that bulged inwards and formed a bowl-like shape with $R^2 > 0.65$ (Fig. S7B). For further details, please refer to the Supplementary Information file.

**Conclusion**

To conclude, this study presented a general framework for investigating the potential final states of liquid marbles based on L–P and P–P interactions. These characteristics can be easily quantified via contact angle goniometry, optical imaging, and forming cones of particles of interest. The behaviors of evaporating liquid marbles formed with particles of sizes varying across four orders of magnitude (7 nm–300 μm) and chemical compositions ranging from hydrophilic to superhydrophobic (Table 1) were carefully studied herein. The behaviors of these diverse liquid marbles can be classified into three categories (Table 2).

- Case I (high L–P adhesion and moderate to low P–P friction): in this case, liquid marbles deform to maintain constant surface area.
- Case II (low L–P adhesion and low P–P friction): in this case, liquid marbles eject particles to maintain sphericity.
- Case III (low L–P adhesion and high P–P friction): in this case, expelled particles remain in the particulate layer and thicken it.

Hybrids of Cases I and III are also observed. Based on these insights, this study developed a model for the evaporation of liquid marbles by exploiting a geometric parameter based on the sphericity of liquid marbles during their time-dependent evolution. This model can accurately capture various aspects of the evaporation of liquid marbles via the three general cases along with the hybrids. These insights should advance the rational understanding of liquid marbles to extend their use in probing complex liquids, such as "dry water"(*61*) and Pickering emulsions (*43, 57, 62-64*) together with their applications as a simple and low-cost platform in engineering and educational contexts.

**Materials and Methods**

Fumed silica particles with diameters of 0.007 and 0.5 µm were obtained from Sigma Aldrich (CAS 112945-52-5). To remove their original coating, they were exposed to three cycles of piranha treatment described below. The resulting surfaces exhibited superhydrophilicity toward water. Then, they were silanized following a protocol described below. Particles in the size range of 3–57 µm were obtained from Davisil (Grade 633–CAS 112926-00-8) with 60 Å pore size (purchased from Sigma-Aldrich). For the particles in the size range of 75–300 µm, silica sand was used, purchased from a rural Saudi supplier. The sand was sieved using stainless steel meshes with ranges specified in Table 1. All silanes (Fig. 1) and chemical reagents (including toluene, chloroform, methanol, sulfuric acid, and hydrogen peroxide) were purchased from Sigma Aldrich. Glaco$^{TM}$ was purchased from Soft99. Deionized Milli-Q water was used for all evaporation experiments, i.e., for water droplets and liquid marbles.

*Silanization of particles*

Silanization reactions were utilized to chemically graft oxide-terminated ceramic/mineral surfaces with molecular species of interest, e.g., hydrocarbons and perfluorocarbons(*65, 66*). First, particles were washed to remove surface contaminants. The particles were stirred in a beaker with acetone for 1 h, then filtered and rinsed with ethanol and dried at 100°C for 6 h in an oven. Subsequently, the particle surface was activated in a freshly made piranha solution (3:1 volume ratio of 99.9% sulfuric acid and 30% hydrogen peroxide) for 10 min at 130°C. The activation step created hydroxyl groups on the surface of particles, which were the binding sites for the silane molecules (Fig. 1A) (*67*). Then, the piranha solution was removed by thoroughly rinsing the particles several times with deionized water. Next, the particles were dried at 100°C for ~2 hours and cooled down. Immediately after, silanization reactions were performed with 1 g of particles and 70 mL of a 1% solution (by volume) of silane in toluene in a stirred beaker at 40°C and 300 rpm for 3 h. The particles were subsequently rinsed in toluene several times to remove the unreacted silanes, oven dried at 100°C, and stored in glass vials. To quantify the coating methodology's effectiveness, a small piece of silicon wafer was introduced into each batch of particles. At the end of the process, the wafer was separated and washed with water and dried in nitrogen to remove any particles from its surface. Then, the surface advancing and receding contact angles were measured in a Kruss Drop Shape Analyzer (DSA100) with deionized water flow rates of 16 µL s$^{-1}$. In few cases, we used Glaco$^{TM}$ to render particles superhydrophobic (S.H.). This coating renders superhydrophobicity by depositing a layer of hydrophobic nanoparticles onto the surface, i.e., the particles are physisorbed. For such cases, Glaco$^{TM}$ was added on the particles that were previously silanized with C18 to guarantee superhydrophobicity. The contact angles for the superhydrophobic case were less homogeneous owing to the variability of this coating protocol, which depends on the thickness of the film deposited on the surface. Values for Glaco$^{TM}$ particles are shown in Table 1 as S.H. The angle of repose was measured by gently pouring the particles over a glass slide and measuring the angle the pile formed with the horizontal plane. For the 7-nm fumed silica particles, the angle of repose exhibited high dependence on interparticle compression. Thus, the reported value for this case is the medium value, and the error represents the difference between the maximum and minimum angles divided by two.

*Preparation of liquid marbles and evaporation experiments*
A layer of the hydrophobic particles was spread onto a glass slide or onto a superhydrophobic paper coated with Glaco™ (S.H-paper—for cases that required more vigorous shaking to attach the particles) and then placed a 10-μL droplet of water on top of the particles. Next, we gently poured more hydrophobic particles until they completely covered the droplet (Fig. 1C). Then, excess particles were swept away with a brush by moving the marble around until only the particles attached to the liquid marble were present. The marble was subsequently transferred to another glass slide (or S.H.-paper, i.e., paper coated with Glaco™), and the process was repeated twice for obtaining a total of three liquid marbles. The glass slide (or S.H-paper) with the three liquid marbles of the same type was placed on a microbalance for evaporation (Mettler Toledo, New Classic MF, ML104/03), which recorded the mass every 1 s. The experiments were performed in a lab-controlled atmosphere at 23 ± 1°C and 60 ± 2% RH. Note that this variance in temperature and humidity was more pronounced between different runs performed on different days while being relatively stable and constant during each evaporation experiment. The evaporation of water droplets was performed by placing them on Glaco™-coated glass slides (H-glass). After complete evaporation of water, the final weight of dry particles was measured to consider the surface particle density, which was normalized by the mass of particles by the initial mass of water (Table 1, Fig. 3). The evaporation time-lapse videos were recorded using a digital microscope (Dino Lite, Edge – AnMo Electronics Corporation).

*Model*
In our model, $D_{\text{char}}$ can be interpreted as the characteristic diameter of a virtual hemispherical water droplet that exhibits the same evaporation rate as the liquid marble. In addition, the normalization of evaporation rate by the initial rate helps eliminate the significantly large contributions of small daily deviations in atmospheric conditions (±1°C and ±2% RH) and exclusively focus on the effects related to marble particles and liquid interactions. In addition, the precision of absolute values calculated using the absolute form of Eq. 10 was considerably enhanced when we considered the saturated vapor concentration, $\rho_{\text{sat}}$, at temperatures between the ambient value (23°C, 60% RH) and that of a wet-bulb (17.7°C). This occurs owing to the evaporative cooling effect, which has been discussed in detail by Kozyrev and Sitnikov (*68*).
The model parameter, $k_e$, was optimized by the minimization of an error function in the form $Error = \sum_i (y_i - f_i)^2$, where $y_i$ is the experimental data point, and $f_i$ is the model point. The curves calculated using our model (Figs. 6E–H) accurately capture our experiments (Figs. 6A–D). This is evidenced by the high values of coefficient of determination, $R^2 = 1 - \sum_i (y_i - f_i)^2 / \sum_i (y_i - \bar{y})^2$, where $\bar{y}$ is the mean of the data points. The error in the prediction of the evaporation rate is higher than that in the mass fraction because the model does not describe the wiggling effect created by particle movement and rearrangement on the interface.
The data used to adjust parameter $k_e$ were obtained during separate runs as the data used to obtain the image fitting data for Fig. S7 to prevent the heat generated by LED lights from increasing temperature during evaporation. However, this does not affect the image data fitting that only correlates the apparent volume of the marbles with the actual mass of liquid. The mass data was smoothened using a quadratic polynomial fit with a moving period of 500 data points (8.3 min.). The evaporation rate was further smoothened with a moving average of 100 data points (1.67 min.). The smoothening parameters were chosen to remove the noise introduced by the microbalance and did not affect the general trend of the data. All data processing was performed in Matlab (R2019a). A list of symbols and abbreviations is provided in Table 3.

**Note:** We chose to normalize the data in a manner that reduced biases in the analysis. Evidently, temperature and relative humidity are crucial variables in determining the concentration of water vapor in air because they govern the magnitude of concentration gradients driving the evaporation. Even though the lab environment is controlled at 23 ± 1°C and 60 ± 2% RH, daily variations are unavoidable. Small variations of 1°C and 2% RH can account for deviations in the initial evaporation rates on the order of 20%, as can be observed in Figs. S4–S5. Therefore, the data must be analyzed in a manner in which the daily variations in such environmental conditions can be neglected. Thus, the evaporation rates were normalized by the initial evaporation rate for each run. Instead of normalizing time by the total evaporation time, the data are viewed in terms of liquid mass fraction ($m/m_o$) to reduce biases toward environmental deviations. Moreover, mass fraction is an easily and precisely measurable parameter, which can be directly connected to the shape of the liquid marble or droplet. In addition, the direction of mass fraction axis was reversed to follow a more intuitive direction of time in the evaporation process.

**Table 3. Symbols and abbreviations**

| Symbol | Meaning |
|---|---|
| $\sigma$ | Surface particle density |
| $A$ | Area of droplet or marble |
| $A_{wet}$ | Particle wet area |
| $P_{wet}$ | Particle wet perimeter |
| $r_{wet}$ | Radius of the particle wetting perimeter |
| $\lambda_C$ | Capillary length |
| $\rho_p$ | Bulk density of particle |
| $\rho_L$ | Bulk density of water |
| $d_p$ | Diameter of particle |
| $F_{int}$ | Interfacial force along the particle wetting perimeter tangential to the liquid surface |
| $F_{adh}$ | Adhesion force between the particle and liquid in the radial direction |
| $F_{fr}$ | Friction force between one particle and its neighboring particles (tangential to P–P) |
| $F_C$ | Compression force experienced by a particle (tangential to liquid) |
| $F_{ej}$ | Ejection force experience by a particle (tangential to P–P) |
| $F_T$ | Resulting total maximum force experienced by a particle (tangential to P–P) |
| $\gamma$ | Liquid–air surface tension of water |
| $R$ | Droplet or marble radius |
| $\theta_a$ | Apparent contact angle at the triple phase interface |
| $\theta_{adv}$ | Advancing contact angle at the triple phase interface |
| $\theta_{rec}$ | Receding contact angle at the triple phase interface |
| $\theta_{hys}$ | Contact angle hysteresis |
| $\theta_{adh}$ | Contact angle at the triple phase interface relative to the P-P tangential direction |
| $\theta_{rep}$ | Angle of repose of particles |
| $\mu$ | Coefficient of friction between particles |
| $j$ | Evaporation flux |
| $j_o$ | Evaporation parameter |
| $\mathfrak{D}_w$ | Diffusion coefficient for water vapor in air |
| $\rho_{sat}$ | Concentration of saturated water vapor |
| $\rho_\infty$ | Concentration of water vapor in the room |
| $\rho_L$ | Liquid density of water |
| $m$ | Mass of water |
| $m_o$ | Initial mass of water |
| $m/m_o$ | Liquid mass fraction |
| $k_e$ | Geometric constant of the evaporation model |
| $D_{char}$ | Characteristic diameter of the marble or droplet |
| $\psi$ | Dissimilarity factor (analogous to sphericity) |
| $V_{L,a}$ | Apparent volume of the liquid front |
| $V_P$ | Volume of particles |
| $V_M$ | Apparent volume of liquid marble |
| $H_M$ | Height of liquid marble |
| $D_M$ | Equatorial diameter of liquid marble |
| $D$ | Diameter |
| $d$ | Base diameter of liquid marble |
| $d_p$ | Particle characteristic dimension |
| $V_{Mo}$ | Initial marble apparent volume |
| RH | Relative humidity |
| NTP | Normal temperature and pressure |
| SH | Superhydrophobic |
| L–P | Liquid–Particle |
| P–P | Particle–Particle |

**Acknowledgments**

AGJ thanks Ms. Jamilya Nauruzbayeva and Mr. Sankara Arunachalam from KAUST for their assistance with the scanning electron microscopy, Ms. Nayara Musskopf (KAUST) for her assistance with the literature review, and Mr. Thiago Reihner and Dr. Meng Shi for the fruitful discussions.

**Funding:** HM acknowledges KAUST for funding (BAS/1/1070-01-01).

**Author contributions:** HM developed the idea. HM and AGJ planned the experiments. AGJ, FT, and RD performed the experiments. AGJ analyzed the data, developed the analytic framework and the evaporation model. AGJ and HM wrote the manuscript and the co-authors provided valuable feedback.

**Competing interests:** The authors declare no competing interests.

**Data and materials availability:** All data needed to evaluate the conclusions in the paper are present in the paper and/or the Supplementary Materials.

61. K. Saleh, L. Forny, P. Guigon, I. Pezron, Dry water: From physico-chemical aspects to process-related parameters. *Chemical Engineering Research and Design* **89**, 537-544 (2011).
62. B. P. Binks, J. H. Clint, Solid wettability from surface energy components: relevance to Pickering emulsions. *Langmuir* **18**, 1270-1273 (2002).
63. B. Liu, W. Wei, X. Qu, Z. Yang, Janus colloids formed by biphasic grafting at a Pickering emulsion interface. *Angewandte Chemie International Edition* **47**, 3973-3975 (2008).
64. S. H. Kim, S. Y. Lee, S. M. Yang, Janus Microspheres for a Highly Flexible and Impregnable Water-Repelling Interface. *Angewandte Chemie International Edition* **49**, 2535-2538 (2010).
65. B. R. Shrestha, S. Pillai, A. Santana, S. H. Donaldson Jr, T. A. Pascal, H. Mishra, Nuclear Quantum Effects in Hydrophobic Nanoconfinement. *The Journal of Physical Chemistry Letters* **10**, 5530-5535 (2019).
66. A. Santana, A. S. F. Farinha, A. Z. Toraño, M. Ibrahim, H. Mishra, A first-principles approach for treating wastewaters. *International Journal of Quantum Chemistry* **121**, e26501 (2021).
67. S. Kaya, P. Rajan, H. Dasari, D. C. Ingram, W. Jadwisienczak, F. Rahman, A systematic study of plasma activation of silicon surfaces for self assembly. *Acs Appl Mater Inter* **7**, 25024-25031 (2015).
68. A. V. Kozyrev, A. G. Sitnikov, Evaporation of a spherical droplet in a moderate-pressure gas. *Physics-Uspekhi* **44**, 725 (2001).
Page **27** of 27

# Supplementary Materials for

## A General Framework for Liquid Marbles

Adair Gallo Jr.[*], Fernanda Tavares, Ratul Das, Himanshu Mishra[*]

*Corresponding author. Email: Himanshu.Mishra@Kaust.edu.sa; Adair.GalloJunior@Kaust.edu.sa

**This PDF file includes:**

Supplementary Text
Figs. S1 to S8
Captions for Movies S1 to S2

**Other Supplementary Materials for this manuscript include the following:**

Movies S1 to S2

## Supplementary Text

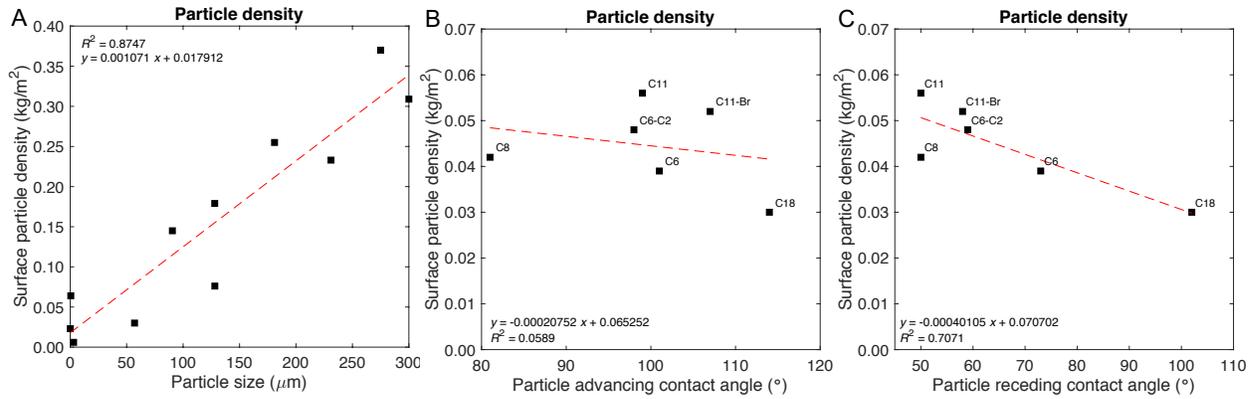

**Fig. S1.**
**Marble surface particle density, $\sigma$, and correlations for 10-µL water marbles. (A)** Effects of varying particle sizes with constant chemical composition (C18). **(B)** Effects of varying chemical compositions of particles at a fixed particle size of 57 µm. Note: particle-level increasing and **(C)** decreasing contact angles of water droplets were obtained via the silanization of flat silica surfaces with the specified functional groups (e.g., C8, C11, etc.).

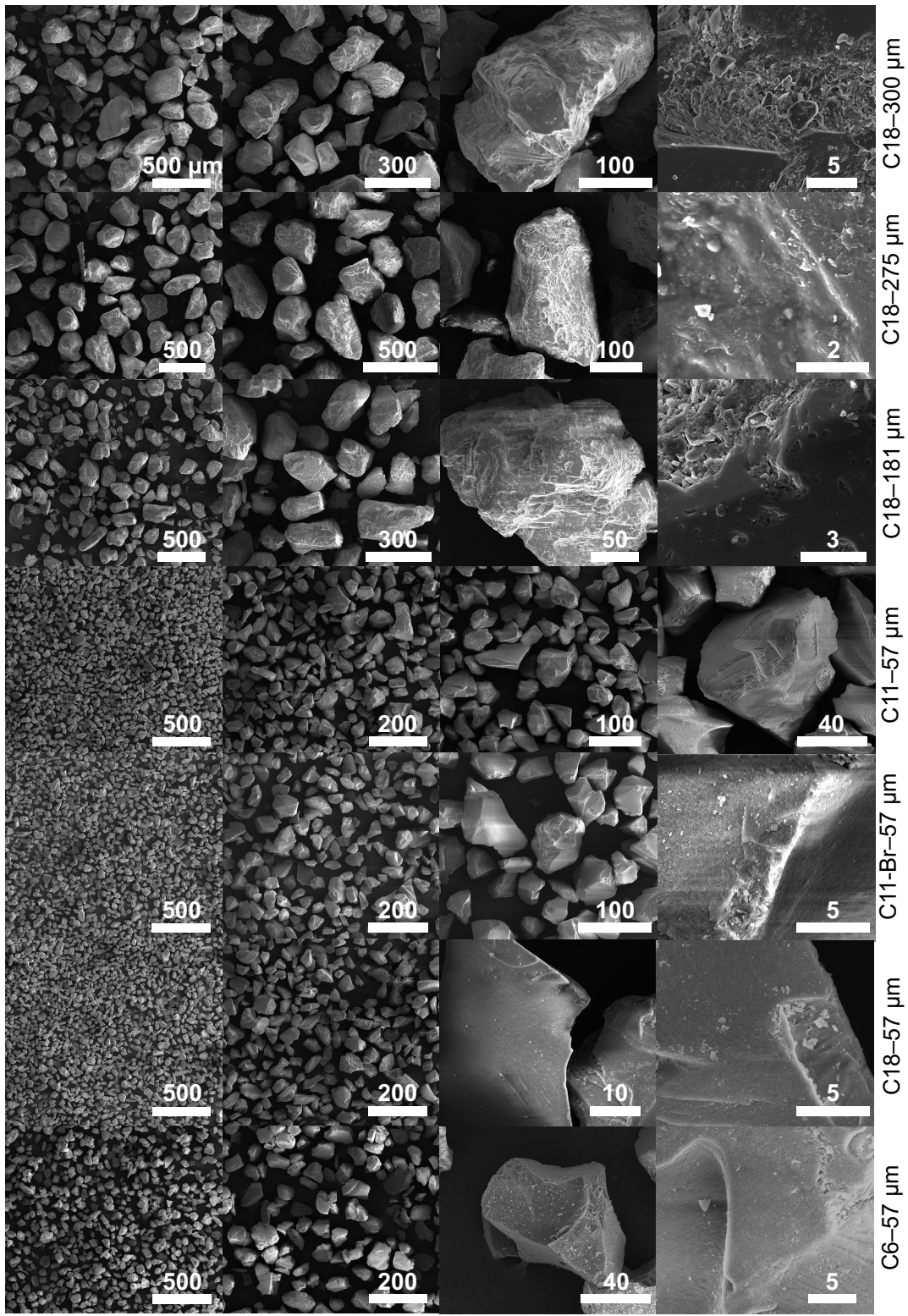

**Fig. S2.**
Scanning electron micrographs of sand and silica particles used to fabricate the liquid marbles. Scale bars are in micrometers.

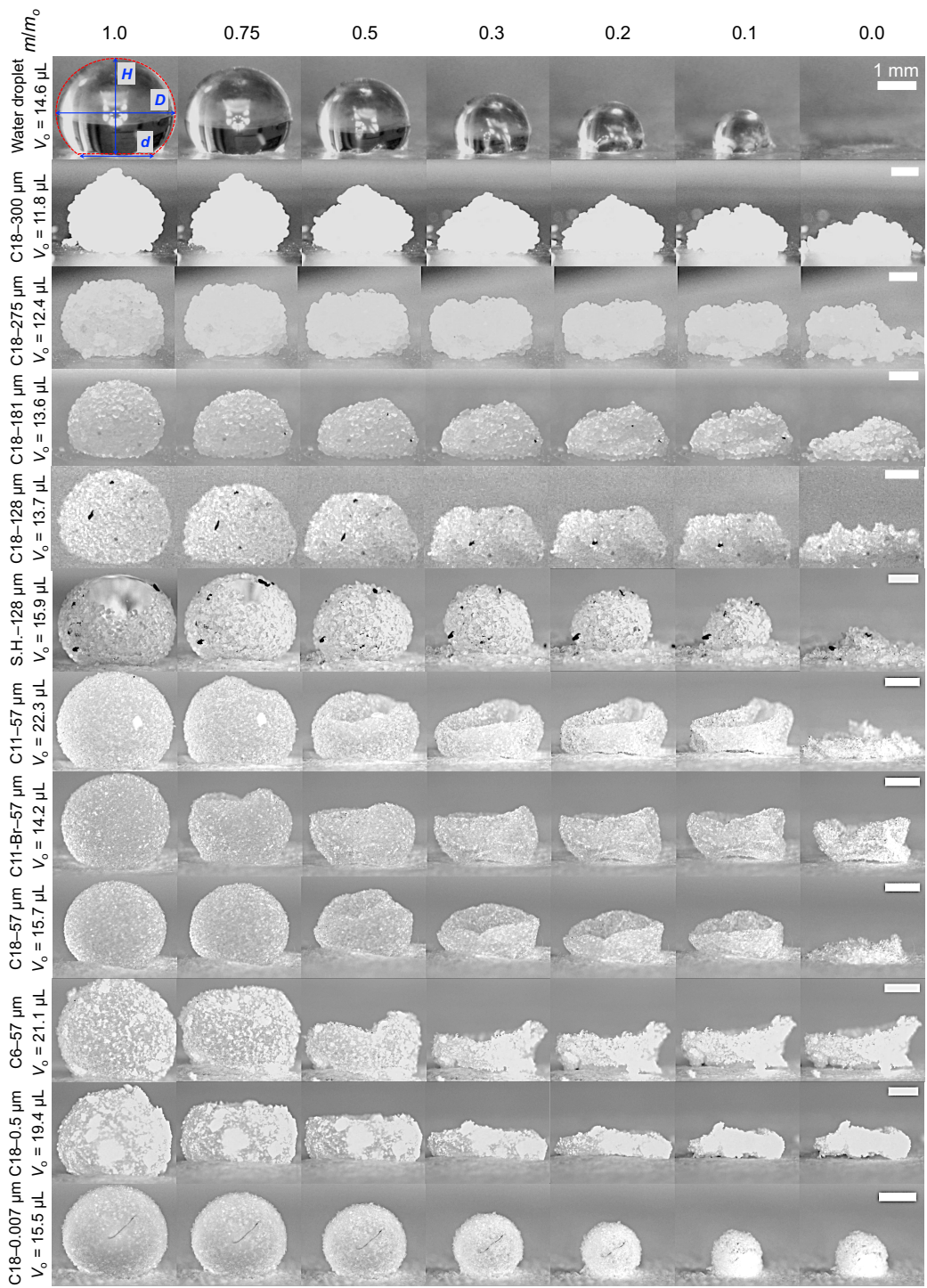

**Fig. S3.**

Time-lapse photographs of the evaporation of (topmost) water droplets on a hydrophobic glass slide (H-glass) and of several water marbles coated with C18; the particle sizes vary from 0.007 to 300 μm.

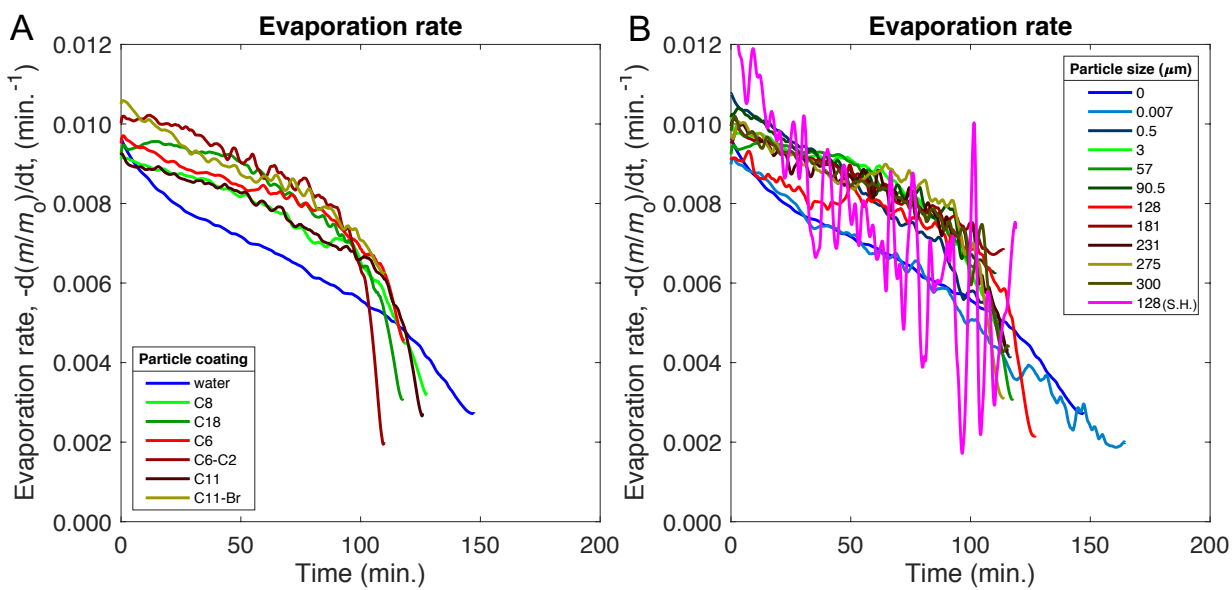

**Fig. S4.**

**Variation in the evaporation rates of liquid marbles over time. (A)** Effects of chemical compositions at a fixed particle size of 57 µm; **(B)** effects of particle size with a fixed chemical composition of C18 (ODTS). Note: The volume of water inside the marbles was 10 µL, and evaporation experiments were performed at 23 ± 1°C and 60 ± 2% relative humidity (RH).

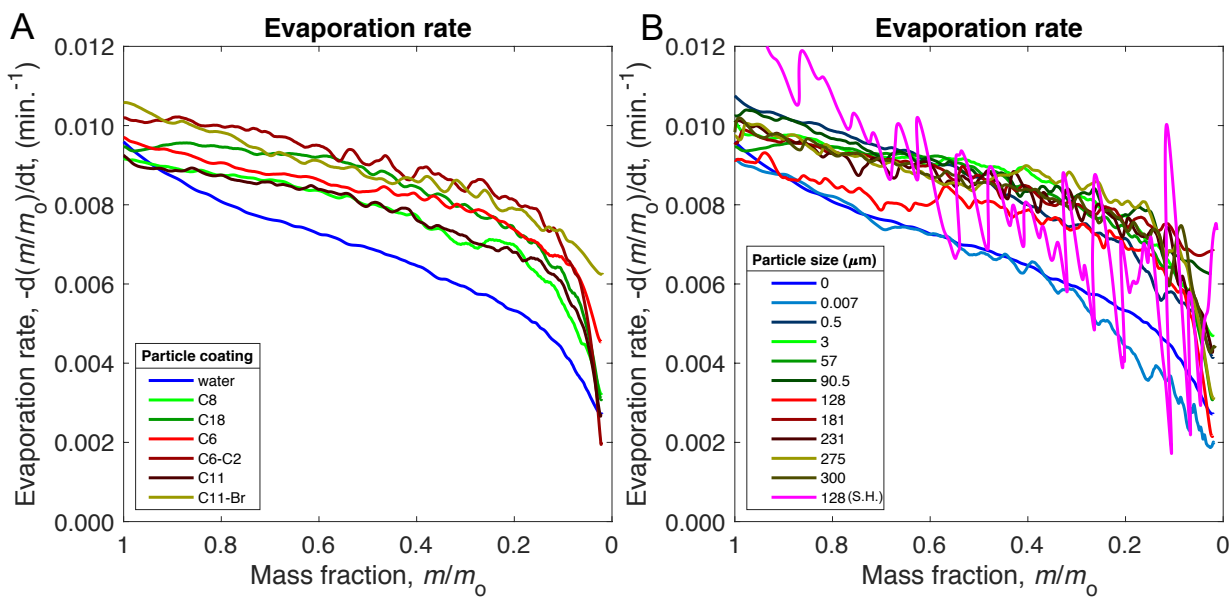

**Fig. S5.**
**Variation in the evaporation rates as a function of mass fraction. (A)** Effects of chemical compositions at a fixed particle size fixed of 57 μm; **(B)** effects of particle size with a fixed chemical composition of C18 (ODTS). Note: The volume of water inside the marbles was 10 μL, and the evaporation experiments were performed at 23 ± 1°C and 60 ± 2% RH.
4

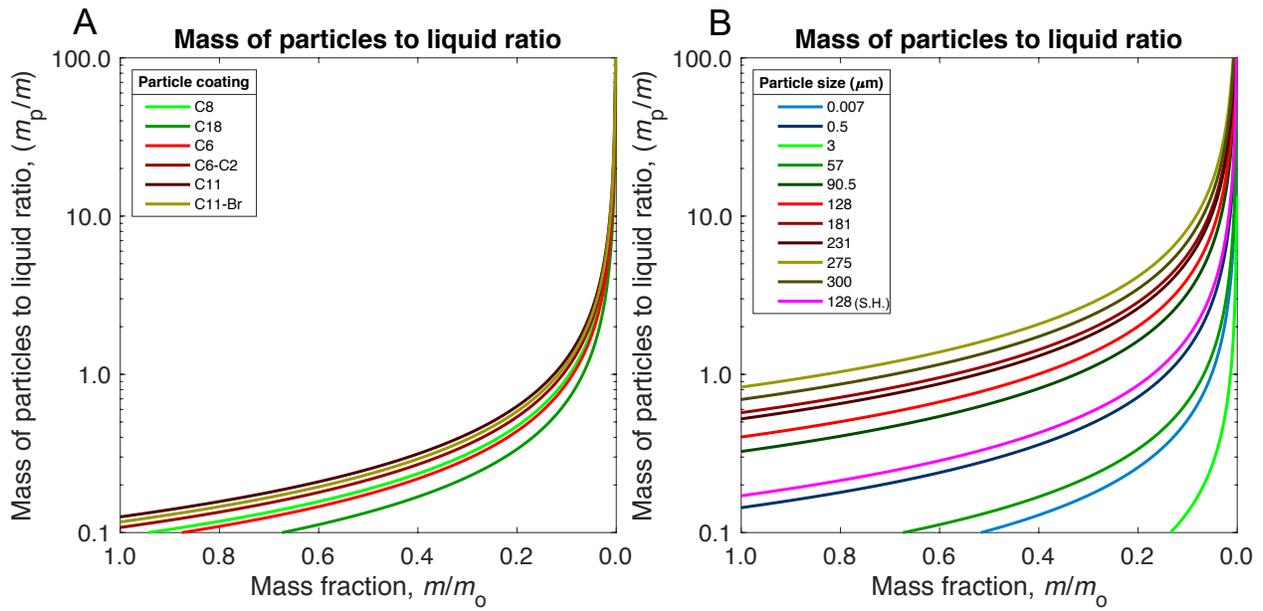

**Fig. S6.**

**Ratio of mass of particles to the mass of liquid for 10-µL liquid marbles. (A)** Different particle coatings at 57-µm particle size, and **(B)** different particle sizes with C18 coating.

Prediction of the apparent volume of liquid marbles

This study derived a formula for the characteristic diameter of liquid in the marble (see Eq. 9 in the manuscript). Accordingly, we can calculate the apparent volume of the liquid front, $V_{L,a}$, approximated as a hemisphere (Eq. S2). This apparent volume of the liquid front (Eq. S1) differs from the actual volume of liquid (measured with a microbalance) because it represents the virtual volume of a hemispherical bare droplet, i.e., it has the same evaporation rate as a hemisphere of diameter $D_{char}$.

$$V_{L,a} = \frac{2}{3}\pi \left(\frac{D_{char}}{2}\right)^3 \quad (S1)$$

By adding the volume of dry particles, $V_P$, the apparent volume of the liquid marble, $V_M$ (Eq. S2), is obtained. The volume of dry particles was experimentally determined through image analysis after complete evaporation of the liquid marbles, followed by logarithmic fitting against the surface particle density (Fig. S7).

$$V_M = V_{L,a} + V_P \quad (S2)$$

To obtain the experimental apparent volume of the liquid marble (Figs. S7B, D), we fitted the time-lapse images with a truncated ellipse of height, $H_M$, equatorial diameter, $D_M$, and base diameter, $d$, adjusted by the correct mass fractions, which were simultaneously recorded by a microbalance. Then, the apparent volumes were calculated through rotation around the central axis to form a truncated ellipsoid. However, note that a simple correlation (Eq. S3) for the measured height and largest diameter of the marble yielded accurate results with less than 3% deviation compared to the values obtained from the truncated ellipsoid. This approach generated greater errors when fitting Case I with smaller particle sizes ($\leq 57\ \mu m$) owing to its more deformed cup-like shape (Fig. 3B). The experimental $D_{char}$ for the water droplet was calculated via volume fitting (Figs. S7A, C).

$$V_M \approx \frac{3\pi}{16} H_M \cdot D_M^2 \quad (Eq.\ S3)$$

Figure S7 shows that the modeled characteristic diameters and apparent volumes were greater than those of the water droplet or that of a reference perfect spherical (or hemispherical) droplet (denoted using the red dashed curve) in all cases but one. The exception was the 7-nm fumed silica marble (only negative $k_e$), which expectedly showed lower $D_{char}$ owing to the increased resistance of vapor diffusion in the particle layer. In addition, evidently the experimental apparent volume of the liquid marble cannot be smaller than that of a spherical droplet; hence, the volume data points for the 7-nm silica marble do not fall below the diagonal curve except owing to fitting error. Notably, the bigger final volumes of the marbles of bigger particles clearly resulted from a higher particle to liquid volumetric fraction, as shown in Fig. S8. Furthermore, all $D_{char}$ go to zero because it is the characteristic diameter of the liquid front, which does not consider the volume of particles.

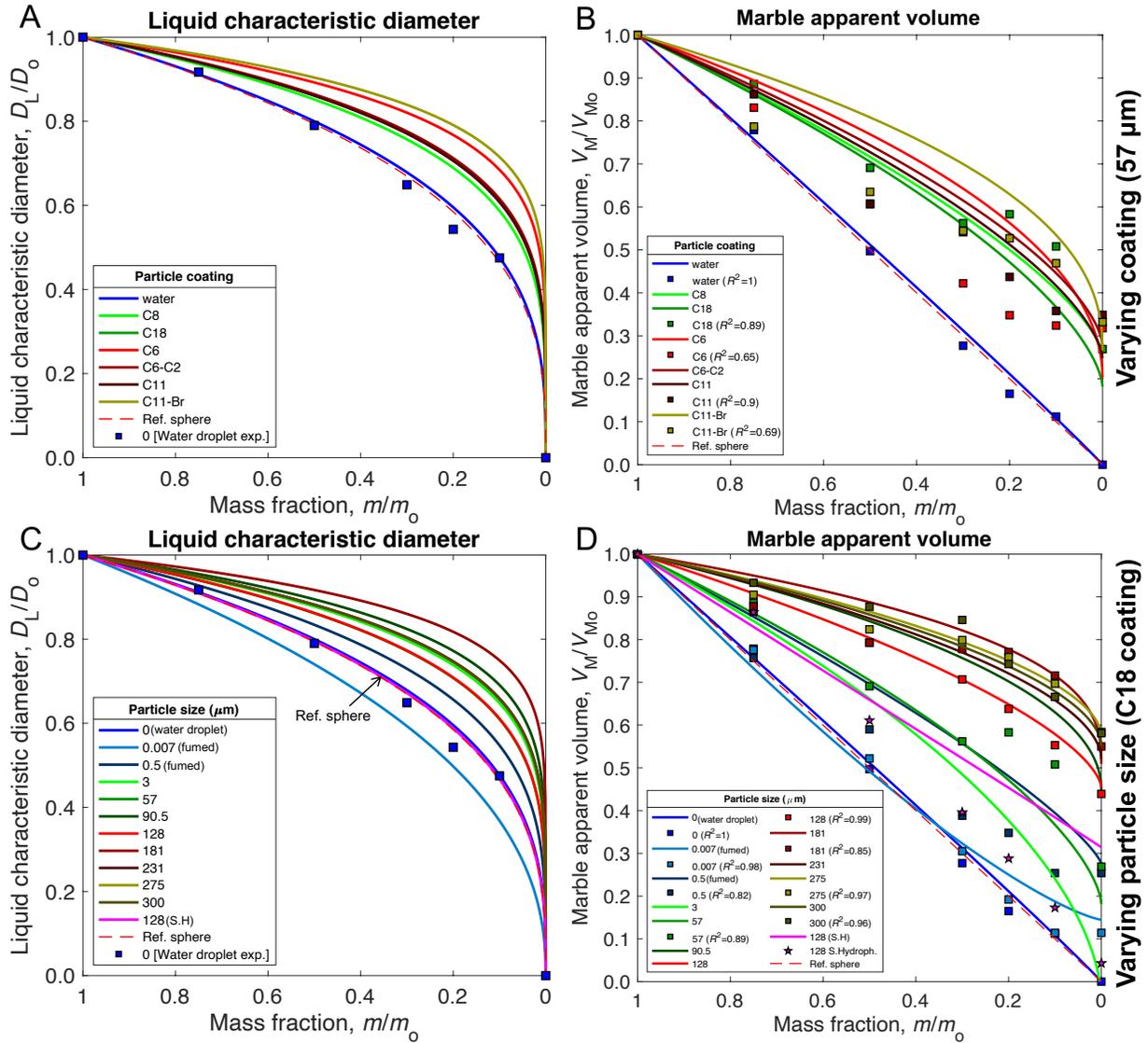

**Fig. S7.**

**Prediction of marble characteristic dimensions.** Experimental values are presented as points, and model predictions are depicted as lines. **(A, C)** Liquid front only and **(B, D)** marble apparent volume compared to experimental results obtained from video analysis. **(A–B)** Varying particle coatings for 57-µm particle size. **(C, D)** Varying particle sizes for C18 coating (ODTS). Experimental data points for the marble apparent volume, $V_M$, are normalized by the initial marble apparent volume, $V_{Mo}$, fitted through image analysis of the evaporation time-lapse videos (Figs. 3 and S2).

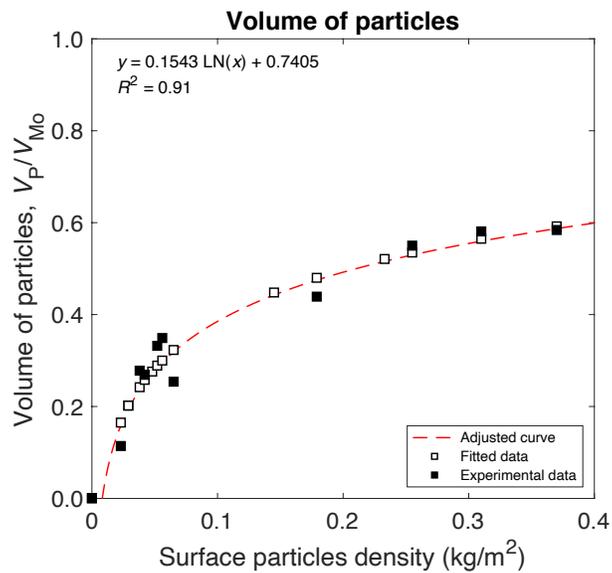

**Fig. S8.**
**Dry volume of particles after complete evaporation.** Volume of particles, $V_P$, relative to the initial volume of the liquid marble, $V_{Mo}$, as a function of the initial surface particle density for 10-µL water marbles.

**Caption for Movie S1.**

High-speed video of superhydrophobic sand grains (C18–128 µm) falling on a water droplet and forming a liquid marble.

**Caption for Movie S2.**

**Time-lapse videos of the evaporation of liquid marbles formed from 10-µL water droplets and particles whose characteristics are described below.** **(A)** Water droplet on a hydrophobic glass slide (H-glass) and without particles (control case). **(B)** Case I for the evaporation of liquid marbles with hydrophobic particles of (Ba) 128 µm and (Bb) 57 µm size, maintaining a monolayer constant surface particle density and marble surface area together with low sphericity. **(C)** Case II for liquid marbles covered with superhydrophobic particles (128 µm) with particle ejection (for $m/m_o < 0.75$) and high liquid sphericity. **(D)** Case III for superhydrophobic fumed particles with increasing surface particle density caused by both increasing thickness and packing of particles at the liquid interface.